\pdfoutput=1

\documentclass[12pt, a4paper]{article}

\usepackage{filecontents}
\usepackage{authblk}

\usepackage{mathtools,amssymb,amsmath,mathdots,amsfonts}
\usepackage{subcaption}
\usepackage{hyperref}
\usepackage{pdfpages}

\usepackage[utf8]{inputenc}
\usepackage[T1]{fontenc}
\usepackage{color}
\newcommand{\tcr}[1]{\textcolor{black}{#1}}
\newcommand{\edit}[1]{\textcolor{black}{#1}}
\newcommand{\add}[1]{\textcolor{black}{#1}}

\makeatletter
\def\input@path{{.},{tex/}}
\makeatother

\usepackage[colorinlistoftodos,bordercolor=orange,backgroundcolor=orange!20,linecolor=orange,textsize=scriptsize]{todonotes}
\usepackage{soul}

\newcommand \eff {*}
\newcommand \scatZ {Z}
\newcommand \scatZs {\zeta}

\newcommand \regS {\mathcal S}

\newcommand \s {\mathbf s}
\newcommand \Ab {\mathcal A}
\newcommand \A [1] {\Ab^#1}
\newcommand \p {p}
\newcommand \prob {P}

\newcommand \reg {\mathcal R_N}
\newcommand \reginf {\mathcal R_\infty}
\newcommand {\nfrac}[1] {\mathfrak n_{#1}}
\newcommand {\Lamo}{\vec \Lambda}
\newcommand {\Lam}[1]{\Lamo_{#1}}

\newcommand \reflect{\mathrm{ref}}
\newcommand \inc{\mathrm{in}}
\newcommand \In{\mathrm{I}}
\newcommand \cs { S}
\newcommand \cl { L}

\newcommand{\ensem}[1]{\langle #1 \rangle}

\newcommand{\be}{\begin{equation}}
\newcommand{\en}{\end{equation}}

\def\bga#1\ega{\begin{gather}#1\end{gather}} 
\def\bgas#1\egas{\begin{gather*}#1\end{gather*}}

\def\bal#1\eal{\begin{align}#1\end{align}} 
\def\bals#1\eals{\begin{align*}#1\end{align*}}

\renewcommand{\vec}[1]{\boldsymbol{#1}}

\newcommand{\ii}{\textrm{i}}
\newcommand{\ee}{\textrm{e}}

\graphicspath{{images/}}

\begin{document}
\title{Reflection from a multi-species material and its transmitted effective wavenumber}

\author[$\dagger$]{Artur L.\ Gower}
\author[$\dagger$]{Michael J.\ A.\ Smith}
\author[$\dagger$]{William J.\ Parnell}
\author[$\star$]{I.\ David Abrahams}

\affil[$\dagger$]{School of Mathematics, University of Manchester, Oxford Road, Manchester M13 9PL, UK}
\affil[$\star$]{Isaac Newton Institute for Mathematical Sciences, 20 Clarkson Road, Cambridge CB3 0EH, UK}

\date{\today}
\maketitle



\begin{center}
random media, size distribution, homogenization.
\end{center}


\begin{abstract}
We formally deduce closed-form expressions for the transmitted effective wavenumber of a material comprising multiple types of inclusions or particles (multi-species), dispersed in a uniform background medium. The expressions, derived here for the first time, are valid for moderate volume fractions and without restriction on the frequency. We show that the multi-species effective wavenumber is not a straightforward extension of expressions for a single species. Comparisons are drawn with state-of-the-art models in acoustics by presenting numerical results for a concrete and a water-oil emulsion in two dimensions. The limit of when one species is much smaller than the other is also discussed and we determine the background \tcr{medium} felt by the larger species in this limit. Surprisingly, we show that the answer is not the intuitive result predicted by self-consistent multiple scattering theories. The derivation presented here applies to the scalar wave equation with cylindrical or spherical inclusions, with any distribution of sizes, densities, and wave speeds. The reflection coefficient \tcr{associated with a half-space of multi-species} cylindrical inclusions is also formally derived.
 \end{abstract}

\section{Summary}
\label{sec:prelude}
Materials comprising mixtures of diverse particles, inclusions, defects or inhomogeneities dispersed inside a background medium arise in a wide range of scenarios and applications, including composite materials, emulsions, gases, polymers, foods, and paints. We will refer to these as multi-species materials.
%
%

Of great importance is the ability to characterise these materials and their {\it microstructure}, such as  particle size distribution and volume fractions. One approach to do this is to employ waves, including electromagnetic, acoustic, and elastodynamic waves.
If either the receivers are much larger than the inclusions, or the wavelength is much longer than the inclusions, then the receivers will measure the \textit{ensemble-averaged} properties of the wave~\cite{pinfield_emergence_2013}. This includes the wave speed, attenuation and reflection. Even methods that estimate fluctuations of the wave on smaller scales, such as the averaged intensity, often require the ensemble-averaged wave properties as a first step~\cite{foldy_multiple_1945,tsang_radiative_1987,tsang_dense_2000}. So in order to improve material characterisation, or to design materials with tailored properties, a crucial step is to rigorously calculate the sound speed and attenuation for multi-species materials.

In this paper, we present and formally deduce the effective wavenumber and reflected field of a plane wave scattered by a material comprising different families, or \textit{species}, of particles with distributions of sizes and properties.
The work here differs from the existing literature as our results are not limited to low frequencies and are valid for moderate number density.
This is achieved by extending the methods introduced in~\cite{linton_multiple_2005} for calculating the effective transmission into a half-space of a single species material.

Our approach does not rely on an extinction theorem or the manipulation of divergent integrals or series. The one assumption that is employed is the {\it quasi-crystalline approximation}~\cite{lax_multiple_1951}.  For a single species, this assumption is supported by theoretical~\cite{martin_multiple_2008,martin_estimating_2010}, numerical~\cite{chekroun_time-domain_2012}
and experimental~\cite{west_comparison_1994} evidence, however the authors are unaware of any rigorous bounds for the error introduced by this assumption. For simplicity, we also restrict attention to the case of circular cylindrical or spherical particles, although our methods can be extended to the case of general-shaped shaped particles by using Waterman's T-matrix approach \cite{waterman_symmetry_1971,varadan_multiple_1978,mishchenko_t-matrix_1996}, for example.




In the context of electromagnetic wave scattering, methods for predicting wave propagation and reflection for multi-species material have previously been developed \cite{tsang_radiative_1987,ding_effective_1988,tsang_dense_2000}. These models have been useful for interpreting data from remote sensing, although it appears that such models cannot systematically reproduce experimental results~\cite{tedesco_intercomparison_2006}. In numerous contexts, but particularly in the context of electromagnetics, the standard approach is to employ the Lippman-Schwinger formulation ~\cite{tourin_multiple_2000,sheng_introduction_2006}. However, such a formulation is restrictive as it is not valid for magnetic media in the electromagnetism context or for scatterers with varying density in acoustics, as identified in \cite{martin_acoustic_2003}. Although it is possible to extend the Lippman-Schwinger formulation to account for this effect~\cite{martin_acoustic_2003}, we found it simpler to extend the multiple scattering theory~\cite{fikioris_multiple_1964,foldy_multiple_1945,linton_multiple_2006}.



Our approach is also in contrast to coupled-phase theory where the first step is to estimate the ensemble average of the governing equations~\cite{spelt_attenuation_2001}, without explicitly considering multiple-scattering. Although this method can accommodate hydrodynamic interactions and has been extended to polydisperse inclusions (multi-species) it does not completely capture multiple-scattering~\cite{spelt_determination_1999,baudoin_extended_2007}.

A suggestion for calculating the multi-species effective wavenumber came from Waterman \& Truell, Equation (3.25a) in the conclusion of \cite{waterman_multiple_1961}. Their suggested formula has been extensively employed in acoustics, see for example \cite{vander_meulen_theoretical_2001,javanaud_experimental_1991,challis_ultrasound_2005}.
 However, their formula is only valid for low frequency and dilute distributions of particles~\cite{linton_multiple_2006}, so it does not properly account for multiple scattering. The approach in  \cite{waterman_multiple_1961} combined with \cite{lloyd_wave_1967} led eventually to the state-of-the-art models for the effective acoustic wavenumber   in colloidal dispersions~\cite{challis_ultrasound_2005}. We numerically compare our results with these authors.

Given an overall particle number density $\nfrac {}$ and background wavenumber  $k$, our main results for a multi-species material comprising \textit{circular cylinders} are the effective transmitted wavenumber:
\begin{equation}
  k_\eff^2 = k^2 - 4 \ii \nfrac {} \ensem{f_\circ}(0) -  4 \ii \nfrac {}^2 \ensem{f_{\circ\circ}}(0)
  + \mathcal O(\nfrac {}^3),
  \label{eqn:wavenumber_results}
\end{equation}
and for an incident wave $u_\inc = \ee^{\ii \alpha x + \ii \beta y}$, with $(\alpha ,\beta) = k(\cos \theta_\inc, \sin \theta_\inc)$, the averaged reflected wave from the inhomogeneous halfspace,
\begin{equation}
 \langle u_\reflect \rangle =  \frac{\nfrac {}}{\alpha^2}\left [
    R_1 + \nfrac {} R_2
  \right]\ee^{-\ii \alpha x + \ii \beta y} + \mathcal O (\nfrac {}^3),
  \label{eqn:reflection_results}
\end{equation}
where
\begin{align}
  & R_1 = \ii \ensem{f_\circ}(\theta_\reflect), \quad \theta_\reflect = \pi - 2 \theta_\inc,
  \label{eqn:R1}
  \\
  & R_2 = \frac{2 \ensem{f_\circ}(0)}{\alpha^2}\left [\frac{\alpha \beta}{k^2}\ensem{f_\circ}'(\theta_\reflect) - \ensem{f_\circ}(\theta_\reflect) \right] + \ii \ensem{f_{\circ\circ}}(\theta_\reflect),
  \label{eqn:R2}
\end{align}
and the functions $\ensem{f_\circ}$ and $\ensem{f_{\circ\circ}}$ are defined in~\eqref{eqns:FarFields} and~\eqref{eqn:MST-pattern}. The formula~\eqref{eqn:reflection_results} is briefly deduced in Section~\ref{eqn:AverageCoefficient}\ref{sec:Reflection},
 and in Figure~\ref{fig:reflect_farfield} we give a pictorial representation, although we stress that the choice $\theta_\reflect = \pi - 2 \theta_\inc$ is not due to a simple geometric argument, but appears from rigorous derivations. From the reflection coefficient~\eqref{eqn:reflection_results} it is possible to choose effective material properties~\cite{caleap_effective_2012}. However, because the reflection coefficient depends on the angle of incidence via $\ensem{f_\circ}(\theta_\reflect)$ and $\ensem{f_{\circ \circ}}(\theta_\reflect)$, it is likely that these effective material properties   change with the angle of incidence.

In the supplementary material we provide a brief self-contained version of these formulas, and the corresponding result for spherical particles, both for a finite number of species. We also provide open source code that implements these formulas, see~\cite{gower_effectivewaves.jl:2017}.
For spherical inclusions the effective transmitted acoustic wavenumber becomes,
\begin{align}
   k_\eff^2 = k^2 -\nfrac {} \frac{4 \pi \ii}{k}  \ensem{F_\circ}(0) + \nfrac {}^2 \frac{(4 \pi)^2 }{k^4} \ensem{F_{\circ\circ}}
   + \mathcal O(\nfrac {}^3),
  \label{eqn:SmallNfracSpheres}
\end{align}
where $\ensem{F_\circ}$ and $\ensem{F_{\circ\circ}}$ are functions associated with scattering from the spherical particle and are defined in~\eqref{eqns:SphericalPatterns}. Note that $\ensem{F_{\circ\circ}}$ has no $\theta$ dependency. For a longer discussion of multiple scattering from spheres see \cite{caleap_effective_2012}.


By developing \textit{multi-species} formulas valid for higher number densities and frequencies, we open up the possibility of characterising and designing a wide  range of advanced materials. The effects of multiple scattering appear only for moderate number density, i.e.\ in the term $\ensem{f_{\circ\circ}}(0)$ in \eqref{eqn:wavenumber_results} and $\ensem{F_{\circ\circ}}$ in \eqref{eqn:SmallNfracSpheres}. One important consequence  of this multiple scattering term  is that a multi-species material can exhibit properties not exhibited by that of a host medium with only one constituent species. We stress that even for just two types of circular cylindrical particles, the effects of multiple scattering are neither intuitive nor easily deduced from the single species case. This becomes apparent in the simple example of a multi-species material where one species is much smaller than the other. In this scenario, we compare our expression for the multi-species effective wavenumber with the state-of-the-art models from acoustics~\cite{challis_ultrasound_2005} and a \textit{self-consistent} type approximation~\cite{sabina_simple_1988,yang_dynamic_2003,kim_generalized_2004}, which can be calculated from the single species formula via an iterative approach: first one \textit{homogenizes} the small particle and background mixture before considering the multiple scattering of the larger particles in the new (homogenized) background medium.
We show analytically that this naive self-consistent methodology is not even correct in the low-frequency limit. This is then demonstrated numerically for the cases of an emulsion and concrete.

The outline of this paper is as follows. In Section~\ref{sec:Multipole} we describe the exact theory of multiple scattering for $N$ cylinders of any radius, density and sound speed. From there we   calculate the effective (ensemble averaged) equations and apply statistical approximations in Section~\ref{sec:Averaged}.  In Section~\ref{sec:EffectiveWavenumbers} we   deduce the governing system for  the effective wavenumbers at arbitrary total number density and arbitrary frequency, before specialising the result to the case of moderate number fraction and low frequency. In Section~\ref{sec:TwoTypes} we investigate the specific, representative case of  two types of circular cylindrical species   and compare different approximations graphically. To calculate the reflected or transmitted wave we also need the effective amplitude, which we calculate in Section~\ref{eqn:AverageCoefficient} followed by the effective reflected wave. We close in Section~\ref{sec:conclusion}, where we discuss avenues for improvement of the techniques and more general further work.



\section{Multipole method for cylinders}
\label{sec:Multipole}

In this section, we describe the exact theory for scalar multiple wave scattering from a finite number $N$ of circular cylinders possessing different densities, \edit{wave} speeds,  and radii. Parameters associated with the medium are summarised in Table~\ref{tab:properties}. Naturally, the system of equations describing this problem bears strong similarities to that obtained by Z\'avi\v ska (see references in \cite{linton_handbook_2001}) and   in \cite{linton_multiple_2005}  for the single species circular cylindrical particle context. Assuming time-harmonic dependence of the form $\ee^{-\ii \omega t}$, the pressure $u$ outside all the cylinders satisfies the scalar Helmholtz equation
\begin{table}[b]
\centering
\begin{tabular}{|l|llll|}
  \hline
Background properties: & &   density $\rho$  & sound speed $c$ &
\\ \hline
Species properties: & number density $\nfrac j$ & density $\rho_j $ & sound speed $c_j$  & radius $a_j$
\\\hline
\end{tabular}
\caption{Summary of material properties and notation. The index $j$ refers to properties of the $j$-th species, see Figure~\ref{fig:multispecies} for an illustration. }
\label{tab:properties}
\end{table}

\begin{figure}[t]
  \centering
  \includegraphics[width=0.72\linewidth]{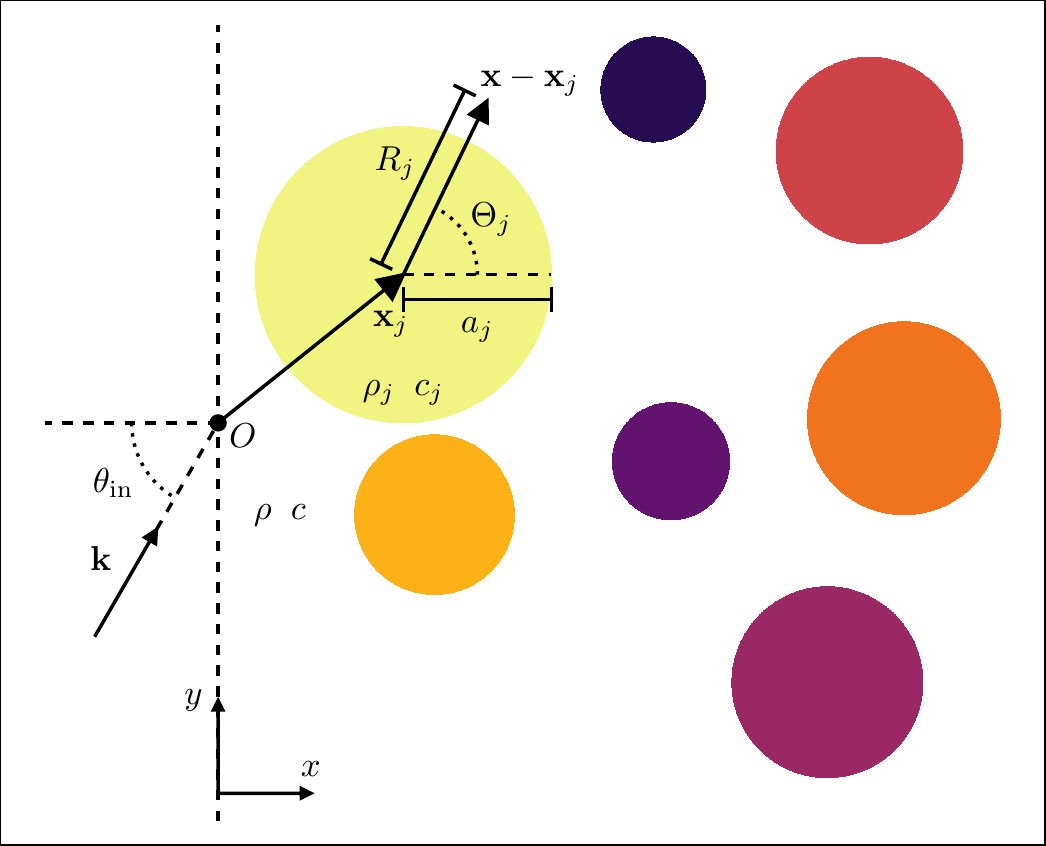}
  \caption{ represents a multi-species material comprising different species of cylinders to the right of the origin $O = (0,0)$. The vector $\mathbf x_j$ points to the centre of the $j$-th cylinder, with a local polar coordinate system $(R_j, \Theta_j)$. Each cylinder has a radius $a_j$, density $\rho_j$, and \edit{wave} speed $c_j$, while the background has density $\rho$ and \edit{wave} speed $c$. The vector $\mathbf k$ is the direction of the incident plane wave.  }
  \label{fig:multispecies}
\end{figure}
\begin{subequations}
\label{eq:governingop}
\be
\nabla^2 u + k^2 u = 0,
\en
and inside the $j$th cylinder the pressure $u_j$ satisfies
\be
  \nabla^2 u_j + k^2_j u_j = 0, \quad \text{for} \; j=1,2,\ldots, N,
\en
\end{subequations}
where $\nabla^2$ is the two-dimensional Laplacian and
\be \label{eqns:wavenumbers}
	k = \omega/c \quad \text{and} \quad k_j = \omega/c_j.
\en
We consider an incident plane wave
\[
u_\inc(x,y) = \ee^{\ii (\alpha x + \beta y)}, \quad \text{with} \;\; (\alpha,\beta) = k ( \cos \theta_\inc,  \sin \theta_\inc),
\]
 and use for each cylinder
the polar coordinates
\be
R_{j} =\| \mathbf x- \mathbf x_{j} \|, \quad \Theta_{j} = \arctan\left ( \frac{y-y_{j}}{x- x_{j}} \right),
\label{eqns:polar_coords}
\en
where $\mathbf x_j$ is the centre of the $j$-th cylinder and $\mathbf x = (x,y)$ is an arbitrary point with origin $O$. See Figure~\ref{fig:multispecies} for a schematic of the material properties and coordinate systems.
Then we can define $u_j$ as the scattered pressure field from the $j$-th cylinder,
\bga \label{eqn:outwaves}
	u_j(R_j,\Theta_j) = \sum_{m=-\infty}^\infty A_j^m \scatZ^m_j H_m(k R_j) \ee^{\ii m \Theta_j}, \quad \text{for} \;\; R_j > a_j,
\ega
where $H_m$ are Hankel functions of the first kind, $A_j^m$ are arbitrary coefficients and $\scatZ^m_j$ characterises the type of scatterer:
\be
\scatZ^m_j = \frac{q_j J_m' (k a_j) J_m (\edit{k_j} a_j) - J_m (k a_j) J_m' (\edit{k_j} a_j) }{q_j H_m '(k a_j) J_m(\edit{k_j} a_j) - H_m(k a_j) J_m '(\edit{k_j} a_j)} = \scatZ^{-m}_j,
\label{eqn:Zm}
\en
with $\edit{q_j} = (\rho_j c_j)/(\rho c)$. In the limits $\edit{q_j} \to 0$ or $\edit{q_j} \to \infty$, the coefficients for Dirichlet or Neumann boundary conditions are recovered, respectively.

The pressure outside all cylinders is the sum of the incident wave $u_\inc$ and all scattered waves,
\be \label{eq:totwave}
u(x,y) =
  u_\inc(x,y) +\sum_{j=1}^N u_j(R_j,\Theta_j),
\en
and the total field inside the $j$-th cylinder is
\be \label{eqn:inwaves}
u_{j}^\In(R_j,\Theta_j) = \sum_{m=-\infty}^\infty B_j^m J_m(k_j R_j) \ee^{\ii m \Theta_j}, \quad \text{for} \;\; R_j < a_j.
\en
In the above, $J_m$ are Bessel functions of the first kind. The arbitrary constants $A_j^m$ and $B_j^m$ in \eqref{eqn:outwaves} and \eqref{eqn:inwaves} will be determined from the boundary conditions of the $j$-th cylinder $R_j = a_j$. The boundary conditions of continuity of pressure and normal   velocity on the cylinder boundaries are given respectively by
\be \label{eqn:BC}
	u = u^\In_{j} \quad \text{and} \quad \frac{1}{\rho} \frac{\partial u}{\partial R_j} = \frac{1}{\rho_j} \frac{\partial u^\In_{j}}{\partial R_j}, \quad \text{on} \;\; R_j = a_j \;\; \text{for} \; \; j=1, \ldots, N,
\en
recalling that $\rho$ and $\rho_j$ denote the material densities of the background and of the $j-$th cylinder respectively. To impose the boundary conditions, we now express the relevant fields in terms of the $(R_j,\Theta_j)$ coordinate system. For the incident wave
\bga
u_\inc(x,y)
 =  I_j \ee^{\ii  k r_j \cos (\theta_j - \theta_\inc)} = I_j \sum_{n = - \infty} ^\infty \ee^{\ii n (\pi/2 - \Theta_j + \theta_\inc)} J_n (k R_j),
\label{eqn:IncidentExpandJ}
\ega
where $I_j = u_\inc(x_j,y_j)$ following  the Jacobi-Anger expansion~\cite{martin_multiple_2006}.
For the scattered waves~\eqref{eqn:outwaves} we use Graf's addition theorem (9.1.79) in \cite{abramowitz1966handbook},
\begin{equation}
  H_n(k R_\ell)\ee^{\ii n \Theta_\ell} =
  \sum_{m=-\infty}^\infty H_{n-m}(k R_{\ell j})\ee^{\ii(n-m)\Theta_{\ell j}} J_m(k R_j)\ee^{\ii m \Theta_j}, \;\;\text{for}\;\; R_j < R_{\ell j},
\label{eqn:Graf}
\end{equation}
where $(R_{\ell j},\Theta_{\ell j})$ is the polar form of the vector $\mathbf x^j-\mathbf x^\ell$. Using the above and \eqref{eqn:IncidentExpandJ} we can impose the boundary conditions~\eqref{eqn:BC} to arrive at the following system of equations

\be \label{eqn:As}
	A^m_j + I_j \ee^{\ii m ( \pi/2 - \theta_\inc )} + \sum_{n=-\infty}^\infty \sum_{\stackrel{\ell =1}{\ell \not= j}}^N A_\ell^n \scatZ^n_\ell \ee^{\ii (n-m) \Theta_{\ell j}} H_{n-m}(k R_{\ell j}) = 0,
\en
for $j=1, \ldots, N$ \edit{and all integers $m$}.
Furthermore, the coefficients associated with the pressure inside the cylinder~\eqref{eqn:inwaves} are then given by
\begin{equation}
  B_j^m = \frac{A_j^m}{J_m(k_j a_j)}\left[ \scatZ_j^m H_m(k a_j) - J_m(k a_j) \right],
\end{equation}
and subsequently the field $u(x,y)$ is entirely prescribed.

In any given material it is impossible to know the exact position and properties of all constituent particles. Our goal is therefore to solve~\eqref{eqn:As} not for one particular configuration of scatterers, but instead to calculate the average value of the coefficients $A^m_j$, denoted by $\langle A^m_j \rangle$, from which we can calculate an effective wavenumber and reflection.
Note that the {\it effective} field describes the {\it ensemble-averaged} field that is   usually   measured in an acoustic experiment, as the receiver face is typically much larger than the particles and the distance between them~\cite{pinfield_simulation_2012,pinfield_emergence_2013}. In our case, we obtain an ensemble-average by  averaging over all particle configurations and all the material properties of the particles. This approach is general and can be tailored to different scenarios, e.g.\ when detailed information \textit{is} known about the particle material properties.

\section{Averaged multiple-scattering}
\label{sec:Averaged}

For an introduction to ensemble-averaging of multiple scattering see~\cite{foldy_multiple_1945} and \cite{parnell_effective_2010}, where the result for a classical dilute isotropic mixture was determined.
Here we present a brief self-contained explanation tailored to multi-species.

Consider a configuration of $N$ circular cylinders centred at $\mathbf x_1,\mathbf x_2, \ldots, \mathbf x_N$ with the scattering properties $\s_1,\s_2, \ldots, \s_N$, where $\s_j$ denotes the properties of the $j$-th cylinder, i.e.\ here these are $\s_j = (a_j,\rho_j, c_j)$. Each $\mathbf x_j$ is in the region $\reg$, where $\nfrac {} = N/|\reg|$ is the total number density and $|\reg|$ is the area of $\reg$. The properties $\s_j$ are taken from the set $\regS$. For example, we could have $\regS = [0,1] \times [1,2] \times [100,200]$, so that $a_j \in [0,1]$,
$\rho_j \in [1,2]$ and $c_j \in [100,200]$.

The probability of the cylinders being in a specific configuration is given by the probability density function $\p(\{\mathbf x_1, \s_1\},\{\mathbf x_2, \s_2\},\ldots, \{\mathbf x_N, \s_N\})$. Using the compact notation $\Lam i = \{\mathbf x_i, \s_i\}$ to denote the properties of the $i$-th cylinder, it follows that
\be
\int \p(\Lam 1) d \Lam 1 = \int \int \p(\Lam 1, \Lam 2) d \Lam 1 d \Lam 2 = \ldots = 1,
\en
where each integral is taken over \textit{both} $\reg$ (for $\mathbf x_j$) and $\regS$ (for $\s_j$) with $d\Lam j = d \mathbf x_i d \s_i$. Note that $\p(\Lam 1, \Lam 2)$ is the probability of one cylinder having the properties $\Lam 1$ and another having the properties $\Lam 2$, when the properties of all the remaining $N-2$ cylinders are unknown. And as the cylinders are indistinguishable: $\p(\Lam 1, \Lam 2) = \p(\Lam 2, \Lam 1)$.
Furthermore, we have
\begin{subequations}
\bal
  &\p(\Lam 1, \ldots, \Lam N) = \p(\Lam j) \p(\Lam {1}, \ldots, \Lam N|\Lam j),
  \label{eqns:conditional_probj}
  \\
  &\p(\Lam 1, \ldots, \Lam N|\Lam j) = \p(\Lam \ell |\Lam j) \p( \Lam 1, \ldots, \Lam N|\Lam \ell,\Lam j),
  \label{eqns:conditional_probsj}
\eal
\end{subequations}
where $\p(\Lam {1}, \ldots, \Lam N|\Lam j)$ is the conditional probability of having cylinders with the  properties $\Lam {1}, \ldots, \Lam N$ (not including $\Lam j$), given that the $j$-th cylinder has the properties $\Lam j$. Likewise, $\p( \Lam 1, \ldots, \Lam N|\Lam{\ell},\Lam{j})$ is the conditional probability of having cylinders with the properties $\Lam {1}, \ldots, \Lam N$ (not including $\Lam \ell$ and $\Lam j$) given that there are already two cylinders present, with properties $\Lam \ell$ and $\Lam j$.

Given some function $F(\Lam{1}, \ldots, \Lam{N})$, we denote its average, or {\it expected value}, by
\be
\ensem F  = \int\ldots \int F(\Lam{1}, \ldots, \Lam{N}) \p(\Lam{1}, \ldots, \Lam{N}) d\Lam{1} \ldots d\Lam{N} .
\en
If we fix the location and properties of the $j$-th cylinder, $\Lam{j}$  and average over all the properties of the other cylinders, we obtain a {\it conditional average} of $F$ given by
\be
\ensem{F}_{\Lam {j}} = \edit{\int\ldots} \int F(\Lam{1}, \ldots, \Lam{N}) \p( \Lam{1}, \ldots, \Lam{N}|\Lam{j}) d  \Lam {1} \ldots \Lam N,
\en
where we do not integrate over $\Lam j$. The average and conditional averages are related by
\bga
\ensem{F}  =   \int  \ensem{F}_{\Lam j} \p(\Lam j) \, d \Lam j \quad \text{and} \quad \ensem{F}_{\Lam j} =  \int  \ensem{F}_{ \Lam j \Lam \ell} \p(\Lam \ell)\, d \Lam \ell,
\label{eqns:conditional_averages}
\ega
where $\ensem{ F}_{\Lam \ell\Lam j}$ is the conditional average when fixing both $\Lam j$ and $\Lam \ell$, and $\ensem{ F}_{\Lam \ell\Lam j} = \ensem{ F}_{\Lam j \Lam \ell}$.

Returning to the task of obtaining effective properties for a multi-species medium, we multiply the system~\eqref{eqn:As} by $\p(\Lam 2,\ldots, \Lam N | \Lam 1)$ and average over $\Lam 2,\ldots, \Lam N$, to reach
\begin{multline*}
  \sum_{n=-\infty}^\infty \sum_{\ell =2}^N \int  \ensem{A_\ell^n}_{\Lam \ell \Lam 1} \scatZ^n(\s_\ell) \ee^{\ii (n-m) \Theta_{\ell 1}} H_{n-m}(k R_{\ell 1}) \p(\Lam \ell | \Lam 1) d \Lam \ell
  \\
  + \ensem{A_1^m}_{\Lam 1} + I_1 \ee^{\ii m ( \pi/2 - \theta_\inc )} =  0,
\end{multline*}
where, without loss of generality, we have chosen $j=1$, used the conditional average definition~\eqref{eqns:conditional_probsj} and defined $\scatZ^n(\s_\ell) := \scatZ_\ell^n$ to make the dependency on $\s_\ell$ explicit. \edit{To further simplify the above, note that
all terms in the sum over $\ell$ give the same value. That is, the terms in the integrand depend on $\ell$ solely through the dummy variable $\Lam \ell$. In particular the probability distribution is the same for each cylinder, and} if $\Lam 2 = \Lam l$, then
$\ensem{A_\ell^n}_{\Lam \ell \Lam 1} = \ensem{A_2^n}_{\Lam 2 \Lam 1}$, \edit{because equation~\eqref{eqn:As} uniquely determines the coefficients $A_\ell^n$ from the position and scattering properties $\Lam \ell$. We use this}
to obtain
\begin{multline}
   \sum_{n=-\infty}^\infty (N - 1) \int  \ensem{A_2^n}_{\Lam 2 \Lam 1}  \scatZ^n(\s_2) \ee^{\ii (n-m) \Theta_{2 1}} H_{n-m}(k R_{21}) \p(\Lam 2 | \Lam 1) d \Lam 2
   \\
  + \ensem{A_1^m}_{\Lam 1} + I_1 \ee^{\ii m ( \pi/2 - \theta_\inc )} = 0.
  \label{eqn:ensemAsystem}
\end{multline}
Our aim is to solve the system above for $\ensem{A_1^m}_{\Lam 1}$, however, this requires that we  make   assumptions about $\p(\Lam 2 | \Lam 1)$ and $\ensem{A_2^n}_{\Lam 2, \Lam 1}$. These approximations are discussed   in   Section~\ref{sec:Averaged}\ref{sec:StatisticalApproximations}, however for the moment, we assume that   an appropriate subsitution has been imposed.

With $\ensem{A_1^m}_{\Lam 1}$, we can calculate the average total pressure (incident plus scattered), measured at some position $\mathbf x$ outside $\reg$, by averaging~\eqref{eq:totwave} to obtain
\be
\ensem{u(x,y)} = u_\inc(x,y) + \sum_{j=1}^N \int \ldots \int u_j(R_j,\Theta_j) \p(\Lam 1, \ldots, \Lam N) d \Lam 1 \ldots d \Lam N,
\en
where $\ensem{u_\inc(x,y)} = u_\inc(x,y)$, because the incident field is independent of the scattering configuration.
We can then rewrite the average outgoing wave $u_j$ by fixing the properties of the $j$-th cylinder $\Lam j$ and using equation~\eqref{eqns:conditional_probj} to reach
\bal
  \ensem{u(x,y)} -u_\inc(x,y) =&  \sum_{j=1}^N \int \ensem{u_j(R_j,\Theta_j)}_{\Lam j} \p(\Lam j) d \Lam j
  =  N \int \ensem{u_1(R_1,\Theta_1)}_{\Lam 1} \p(\Lam 1) d \Lam 1.
\label{eqn:AverageWave}
\eal
Likewise, for the conditionally averaged scattered field~\eqref{eqn:outwaves} measured at $\mathbf x$ we obtain
\bga
	\ensem{u_1(R_1,\Theta_1)}_{\Lam 1} = \sum_{m=-\infty}^\infty \ensem{A_1^m}_{\Lam 1} \scatZ^m(\s_1) H_m^{(1)}(k R_1) \ee^{\ii m \Theta_1}.
\label{eqn:AverageWaveCond}
\ega
We use the above to calculate the reflection from a halfspace in Section~\ref{eqn:AverageCoefficient}\ref{sec:Reflection} and to obtain~\eqref{eqn:reflection_results}. To proceed we need to solve the system~\eqref{eqn:ensemAsystem} and, in line with existing approaches, we do this by making statistical approximations.

\subsection{Statistical approximations}
\label{sec:StatisticalApproximations}
In order to solve \eqref{eqn:ensemAsystem} for $\ensem{A_1^n}_{\Lam 1}$ we need an approximation for $\ensem{A_2^n}_{\Lam 2, \Lam 1}$ and the pair distribution $p(\Lam 2 | \Lam 1)$.
In this work, we adopt the standard \emph{closure} approximation for single species, but extended to multi-species, the {\it quasicrystalline approximation} (QCA) \cite{lax_multiple_1951,linton_multiple_2005}:
\be
\ensem{A_2^n}_{\Lam 2 \Lam 1} \approx \ensem{A_2^n}_{\Lam 2}.
\label{eqn:QCA}
\en
 This approximation still makes sense for multi-species because it replaces the dependence of $\ensem{A_2^n}_{\Lam 2 ,\Lam 1}$ in $\Lam 1$ by its expected value in $\Lam 1$. Note also that the expected difference in $\Lam 2$:
\[
\int \left[ \ensem{A_2^n}_{\Lam 2 \Lam 1} - \ensem{A_2^n}_{\Lam 2}\right] p(\Lam 2) d \Lam 2 =
\ensem{A_2^n}_{\Lam 1} - \ensem{A_2^n} \approx 0,
\]
for a large number of scatterers.

Using QCA, we introduce the notation
\be
 \mathcal A^n  (\mathbf x_j, \s_j) = \ensem{A_j^n}_{\Lam j}, \quad \text{and} \quad \mathcal A^n  (\mathbf x_j, \s_j) =  \ensem{A_j^n}_{\Lam j \Lam k} \quad \text{for} \quad k \not = j.
\label{eqn:Afunc}
\en

Next, we determine a suitable approximation for the pair distribution $p(\Lam 2 | \Lam 1)$, beginning with \eqref{eqns:conditional_probj} to write
\begin{equation}
\label{eqn:pcond21}
p(\Lam 2 | \Lam 1) = \left[ p(\Lam 1) \right]^{-1}  p(\Lam 1,\Lam 2).
\end{equation}
For clarity, we introduce the spatial random variables $\mathbf X_1, \, \mathbf X_1, \ldots, \mathbf X_N$ and the scattering property random variables $\mathbf S_1, \,\mathbf S_1, \ldots, \mathbf S_N$, and write  probability density functions in the form
\be
\p(\Lam 1, \ldots, \Lam N) = \prob (\mathbf X_1 = \mathbf x_1, \ldots, \mathbf X_N = \mathbf x_N, \mathbf S_1 = \s_1, \ldots,\mathbf S_N = \s_N),
\en
for example.
In the first instance, we assume the random uniform distribution
\be
\p(\Lam 1) = \frac{1}{|\reg|} \prob(\mathbf S_1 = \s_1),
\label{eqn:pLam1}
\en
where $\prob(\mathbf S_1 = \s_1)$ is the probability density in $\regS$ that the particle will have scattering property $\s_1$. The above assumes that $\prob(\mathbf X_1 = \mathbf x_1 |\mathbf S_1 = \s_1) = |\reg|^{-1}$, i.e.\ that the position $\mathbf x_1$ of the cylinder is independent of the scattering property $\s_1$. This is not always the case, for example, depending on the size of the cylinder, some positions near the boundary of $\reg$ may be infeasible. However, these boundary effects are negligible when taking the limit $|\reg| \to \infty$.

For the remaining distribution in \eqref{eqn:pcond21} we use
\be
\label{eqn:PconditionalS}
\p(\Lam 1, \Lam 2) = \prob(\mathbf S_1 = \s_1, \mathbf S_2 = \s_2) \prob(\mathbf X_1 = \mathbf x_1, \mathbf X_2 = \mathbf x_2 | \mathbf S_1 = \s_1, \mathbf S_2 = \s_2),
\en
followed by
\be
\label{eqn:partitionSprob}
\prob(\mathbf S_1 = \s_1, \mathbf S_2 = \s_2) = \prob(\mathbf S_1 = \s_1)\prob(\mathbf S_2 = \s_2),
\en
which assumes that the scattering properties are statistically independent.
Next we assume that the cylinders are equally likely to be anywhere but do not overlap (a \emph{hole correction} correlation~\cite{fikioris_multiple_1964}), to write
\be
\prob(\mathbf X_1 = \mathbf x_1,\mathbf X_2 =\mathbf x_2 | \mathbf S_1 = \s_1, \mathbf S_2 = \s_2) =
  \begin{cases}
   0  &\text{if} \;\; R_{21} \leq a_{21},   \\
  |\reg|^{-2}  &\text{if} \;\; R_{21} > a_{21},
  \end{cases}
\label{eqn:PcondS}
\en
 where $ R_{2 1} := \|\mathbf x_1 - \mathbf x_2\|$, $a_{21} =  b_1+b_2$ for some $b_{1} \geq a_1$ and some $b_2 \geq a_2$, and $b_1$ is the radius of exclusion around $\mathbf x_1$ which is usually chosen to be proportional to the radius $a_1$.  Note that when integrating \eqref{eqn:PcondS} above in $\mathbf x_1$ and $\mathbf x_2$ we obtain $|\reg|^{-2} (|\reg|^2 - \pi a_{21}^2 ) \approx 1$ for $\reg \gg a_{21}^2$.

 Ultimately, substituting \eqref{eqn:partitionSprob} and \eqref{eqn:PcondS} into \eqref{eqn:PconditionalS} in tandem with \eqref{eqn:pLam1}
 leads to the pair distribution
\bal
\p(\Lam 2 | \Lam 1)  
 =& \frac{1}{|\reg|}\p(\s_2) H(R_{21}-a_{21}),
\label{eqn:PcondSimplified}
\eal
where $H(x)$ denotes the Heaviside function, under  the assumption $|\reg| \gg a_{21}^2$.
 In the next section we will the approximations \eqref{eqn:Afunc} and \eqref{eqn:PcondSimplified} to solve the system in \eqref{eqn:ensemAsystem} for $\ensem{A_1^n}_{\Lam 1}$.

We now include a discussion of other commonly used pair distributions.
We remark that for densely packed scatterers, other pair distributions~\cite{caleap_coherent_2012} are preferred and take the form
\be
\prob(\mathbf X_1 = \mathbf x_1,\mathbf X_2 =\mathbf x_2 | \mathbf S_1 = \s_1, \mathbf S_2 = \s_2) =
  \begin{cases}
   0 & \text{if} \;\; R_{21} \leq a_{21},   \\
   \frac{1 + \chi(R_{21} | \mathbf s_1, \mathbf s_2)}{|\reg|^2}  &\text{if} \;\; R_{21} > a_{21},
  \end{cases}
\label{eqn:PcondSGeneral}
\en
where
\begin{equation}
  \int_{\reg} \int_{\reg} \chi(R_{21} | \mathbf s_1, \mathbf s_2) d \mathbf x_1 d \mathbf x_2 = 0.
  \label{eqn:ChiRestriction}
\end{equation}
To calculate the effective wavenumber for the pair-correlation~\eqref{eqn:PcondSGeneral}, a common choice is to assume that the scatterers are uniformly randomly distributed, which leads to
\begin{equation}
 \chi(R_{21} | \mathbf s_1, \mathbf s_2) \approx 0 \quad \text{for} \quad R_{21}> \bar a_{21} > 2 a_{21},
 \label{eqn:pairdist_assumption}
\end{equation}
 used by \cite[Section D]{linton_multiple_2005}, \cite{bose_ultrasonic_1996} and \cite[Eq. (27)]{twersky_acoustic_1978}, where $\bar a_{21}$ is some distance large enough for the scatterers at $\mathbf x_1$ and $\mathbf x_2$ to no longer effect each other.
One popular choice for $\chi$ is the Percus--Yevick function, which assumes all scatterers are uniformly randomly distributed~\cite{ding_monte_1992}, though $\chi$ can also be used to specify if some species are more likely to be closer or further apart.

In this work we set $\chi = 0$ for simplicity (unless otherwise stated), but also because it is not clear that the error introduced by using $\chi = 0$ is in any way greater than the error committed due to QCA~\eqref{eqn:QCA}. Both the hole correction~\eqref{eqn:PcondS} and QCA \eqref{eqn:QCA} make similar assumptions: for $R_{21} > a_{21}$, the hole correction replaces $p(\Lam 2 | \Lam 1)$ with its expected value in $\Lam 1$:
\[
p(\Lam 2 | \Lam 1) \approx \int p(\Lam 2 | \Lam 1) p(\Lam 1) d \Lam 1 =  p(\Lam 2),
\]
just as QCA~\eqref{eqn:QCA} assumes that $\ensem{A^n_2}_{\Lam 2 \Lam 1} \approx \ensem{A^n_2}_{\Lam 2}$. Similarly for $R_{21} \leq a_{21}$ we would set both $p(\Lam 2 | \Lam 1) = 0$ and $\ensem{A^n_2}_{\Lam 2 \Lam 1} = 0$ for QCA and hole correction.
Another reason to set $\chi = 0$ is because we are interested in the limit for small $\nfrac {}$. In this limit, it is expected that $\chi \to 0$ when $\nfrac {} \to 0$ for uniformly distributed scatterers, which in turn indicates that the contribution of $\chi$ to the effective wave is smaller than $\nfrac {}^2$ \cite{bose_elastic_1974,linton_multiple_2005}.
%
%

\subsection{Infinitely many cylinders in the half space}

In   preceding sections,  we   considered a finite number of scatterers in a bounded domain $\reg$. Now we   consider the limit $N \to \infty$ and where the region $\reg$ tends to the halfspace $x>0$. We will follow~\cite{linton_multiple_2005} and limit the cylinders to the halfspace $x>0$, as it allows us to avoid divergent integrals, such as those in \cite{bose_longitudinal_1973}, e.g. their equation between (32) and (33).

Substituting the approximations \eqref{eqn:PcondSimplified} and \eqref{eqn:Afunc} into the governing system~\eqref{eqn:ensemAsystem} leads to
\begin{multline}
\frac{N-1}{|\reg|} \sum_{n=-\infty}^\infty \int_\regS \int_{\stackrel{\reg}{R_{2 1} > a_{21}}}  \A n (\mathbf x_2, \s_2) \ee^{\ii (n-m) \Theta_{21}} H_{n-m}(k R_{21})  d \mathbf x_2 d\s_2^n
\\
+  \A m (\mathbf x_1, \s_1) + I_1 \ee^{\ii m ( \pi/2 - \theta_\inc )}
   = 0, \quad \text{for} \quad x_1 >0,
  \label{eq:prelim}
\end{multline}
where for brevity we write
\begin{equation}
  d\s_2^n = \scatZ^n (\s_2)\p(\s_2) d\s_2,
  \label{eqn:ds2}
\end{equation}
with $\p(\s_2) = \prob(\mathbf{S}_2 = \s_2)$.
By taking the limits $N\to \infty$ and $\lim_{N \to \infty}\limits \reg = \left\{ (x_1,x_2): x_2>0\right\}$, while fixing the number density $\nfrac {} = \frac{N}{|\reg|}$, equation~\eqref{eq:prelim} takes the form
\begin{multline}
\nfrac {} \sum_{n=-\infty}^\infty \int_\regS \int_{\stackrel{x_2 > 0 }{R_{2 1} > a_{21}}}  \A n (\mathbf x_2, \s_2) \ee^{\ii (n-m) \Theta_{21}} H_{n-m}(k R_{21})  d \mathbf x_2 d\s_2^n
\\
+  \A m (\mathbf x_1, \s_1) + I_1 \ee^{\ii m ( \pi/2 - \theta_\inc )}
   = 0, \quad \text{for} \quad x_1 >0,
  \label{eqn:ensemAsystem2}
\end{multline}
which represents the governing system for our semi-infinite multiple-species problem.

Incidentally, when all cylinders are identical this system reduces to equation (54) in \cite{linton_multiple_2005}, that is when $\p(\s_2) = \delta(a_2 - a ) \delta(c_2 -   c ) \delta(\rho_2 -   \rho )$ in \eqref{eqn:ds2}, where $\delta (x)$ represents Dirac's delta function.

\section{Effective wavenumber}
\label{sec:EffectiveWavenumbers}
%
To solve the system~\eqref{eqn:ensemAsystem2} first we use the symmetry of the problem to rewrite
\begin{equation}
  \A m(x ,y, \s) =   \A m(x, 0, \s) \ee^{\ii \beta y  },
\label{eqn:AnsatzA1}
\end{equation}
that is, if $\A m$ is a solution to \eqref{eqn:ensemAsystem2}, then so is $\A m_0$ defined by $\A m_0(x ,y, \s) = \A m(x ,y-y', \s)\ee^{\ii \beta y'}$ for every $y'$, then taking $y'=y$ we see that \eqref{eqn:AnsatzA1} is also a solution, recalling that $I_1 = \ee^{\ii \alpha x + \ii \beta y}$ and
\begin{equation}
\label{eq:incwavecosines}
  \alpha = k \cos \theta_\inc \quad \text{and} \quad \beta = k \sin \theta_\inc.
\end{equation}
Sufficiently far away from the boundary, say $x > {\bar x}$, we assume a plane wave ansatz
\begin{align}
  \A m(x ,y, \s) =   \ii^m \ee^{-\ii m \theta_\eff} \A m_\eff (\s) \ee^{\ii \mathbf k_\eff \cdot \mathbf x}, \quad \text{for} \quad x>{\bar x},
  \label{eqn:AnsatzA}
\end{align}
where the factor $\ii^m \ee^{-\ii m \theta_\eff}$ is introduced for later convenience.
We could have for generality considered a sum of plane waves, but for low number density this is unnecessary, as we would find a unique $\mathbf k_\eff$ for a halfspace.

Equating \eqref{eqn:AnsatzA1} and~\eqref{eqn:AnsatzA}, for $x > \bar x$, we obtain Snell's law
\begin{equation}
  k_\eff \sin \theta_\eff = k \sin \theta_\inc  \quad \text{with} \quad \mathbf k_\eff =(\alpha_\eff, \beta) := k_\eff(\cos\theta_\eff, \sin\theta_\eff),
  \label{eqn:Snells}
\end{equation}
  noting that both $\theta_\eff$ and $k_\eff$ are complex numbers. We also require that $\mathrm{Im } \,  \alpha_\eff > 0$, so that the integral over $\mathbf x_2$ in~\eqref{eqn:ensemAsystem2} converges. 

In Appendix~\ref{sec:Integrals} we present the derivation for the system below, which is obtained by substituting~\eqref{eqn:AnsatzA1} and~\eqref{eqn:AnsatzA} into \eqref{eqn:ensemAsystem2}. In the process we establish that $k_\eff \neq k$, and find that there is no restriction on the length $\bar x$, a fact we use to calculate the reflected wave. The result is that~\eqref{eqn:ensemAsystem2} reduces to the system
\begin{align}
    & \A m_\eff(\s_1)  +  2 \nfrac {} \pi \sum_{n=-\infty}^\infty\int_\regS  \A n_\eff(\s_2)
  \left [ \frac{\mathcal N_{n-m}(ka_{12},k_\eff a_{12})}{k^2 - k_\eff^2}  +  \mathcal X_{\eff} \right]
    d\s_2^n
   = 0,
 \label{eqn:AmT}
\\
  &
   \sum_{n=-\infty}^\infty \ee^{\ii n (\theta_\inc - \theta_\eff)} \int_\regS
   \A n_\eff(\s_2) d\s_2^n = \ee^{\ii (\alpha-\alpha_\eff){\bar x}}(\alpha_\eff-\alpha) \left [\frac{\alpha \ii}{2 \nfrac {}} +  b(\bar x)\right],
 \label{eqn:AmInc}
 \end{align}
in terms of the unknown parameters $\A n_\eff(\s_2)$ and $k_\eff$, where
\begin{align}
  & b(\bar x) =  (-\ii)^{n-1} \sum_{n=-\infty}^\infty \ee^{\ii n \theta_\inc} \int_\regS
     \int_{0}^{{\bar x}} \A n(x_2,0, \s_2)  \ee^{-\ii \alpha x_2}d  x_2  d\s_2^n,
 \label{eqn:bbar} \\
& \mathcal N_n(x,y) = x H_n'(x) J_n(y) - y H_n(x) J_n'(y),
\label{eqn:Nn}
\end{align}
and $\mathcal X_{\eff} = 0$, as we have assumed {\it hole correction}~\eqref{eqn:PcondS}. For a more general pair distribution~\eqref{eqn:PcondSGeneral}, we obtain
\begin{equation}
   \mathcal X_{\eff} =  \int_{a_{21}< R < \bar a_{21}} H_{n-m}(k R) J_{n-m}(k_\eff R) \chi(R | \mathbf s_1, \mathbf s_2)  R \, d R,
   \label{eqn:X_eff}
\end{equation}
where further    details   may be found in Appendix~\ref{sec:Integrals}\ref{sec:PairDistribution}. We also remark that  equation \eqref{eqn:AmT} reduces to  \cite[Eq. (33)]{bose_longitudinal_1973} and   \cite[Eq. (87)]{linton_multiple_2005} for a single particle species.

To determine the effective wavenumber $k_\eff$ we need only use~\eqref{eqn:AmT}. That is, the solution $k_\eff$  is the one that leads to non-trivial solutions for the function $\A m_\eff$.
On the other hand, if $\A m_\eff(\s_1)$ is a solution to \eqref{eqn:AmT}, then so is $c \A m_\eff(\s_1)$ for any constant $c$. To completely determine $\A m_\eff(\s_1)$ we need to use \eqref{eqn:AmT} and \eqref{eqn:AmInc}.

Next, we determine closed-form estimates for $k_\eff$ from \eqref{eqn:AmT}, and determine the corresponding coefficients $\A m_\eff(\s_1)$  for low number density in Section~\ref{eqn:AverageCoefficient}.

\subsection{Explicit expressions for $k_\eff$ via expansions in the number density}

We now consider the expansions
\begin{align}
  & k_\eff^2 \sim K_{\eff 0} + K_{\eff 1} \nfrac {} + K_{\eff 2} \nfrac {}^2 \quad \text{and} \quad \A m_\eff\sim \A m_{\eff 0} + \A m_{\eff 1} \nfrac {} + \A m_{\eff 2} \nfrac {}^2,
  \label{eqns:nfracExpansion}
\end{align}
where we use $\sim$ to denote an asymptotic expansion in $\nfrac {}$, which is formally equivalent to an expansion in   volume fraction. We show in Appendix~\ref{sec:Integrals}\ref{sec:app_LowNumberFraction} how substituting the above into equation~\eqref{eqn:AmT} leads to
\begin{align}
   k_\eff^2 = k^2 - 4 \nfrac {} \ii \ensem{f_\circ}(0) - 4 \nfrac {}^2 \ii \ensem{f_{\circ \circ}}(0) + \mathcal O(\nfrac {}^3),
  \label{eqn:SmallNfrac}
\end{align}
where we assumed $K_{\eff 0} = k^2$
, though we deduce this in Section~\ref{eqn:AverageCoefficient}. Here we have
\begin{equation}
 f_\circ(\theta,\s_1) = - \sum_{n=-\infty}^\infty \ee^{\ii n \theta} \scatZ^n(\s_1) \quad \text{and} \quad
 \ensem{f_\circ}(\theta) = - \sum_{n=-\infty}^\infty \ee^{\ii n \theta} \int_\regS  d\s_1^n,
 \label{eqns:FarFields}
\end{equation}
and we introduce the multiple-scattering pattern
\begin{equation}
  \ensem{f_{\circ \circ}}(\theta) = - \pi \sum_{n,m=-\infty}^\infty \int_\regS \ee^{\ii n \theta} a_{12}^2 d_{n-m}(k a_{12}) d\s_1^n d\s_2^m,
  \label{eqn:MST-pattern}
\end{equation}
where  $d\s_1^n = \scatZ^n(\s_1) p(\s_1) d\s_1$ and for convenience we   define
\begin{equation}
  d_m(x) = J_m'(x) H_m'(x) + (1 - (m/x)^2)J_m(x) H_m(x).
  \label{eqn:d}
\end{equation}
We remark that  $f_\circ$ corresponds to the far-field scattering pattern for a single circular cylinder. This is evident by taking  $N=1$ in \eqref{eqn:As}, which leads to
\begin{equation}
 A^m_1 = -\ii^m \ee^{-\ii m \theta_\inc} \ee^{\ii \mathbf x_1 \cdot \mathbf k},
 \label{eqn:SingleScatterer}
\end{equation}
and from~\eqref{eqn:outwaves} gives
\begin{equation*}
 \lim_{R_1 \to \infty }u_1 \sim
 \sqrt{\frac{2}{\pi k R_1}} f_\circ(\Theta_1-\theta_\inc,\s_1)\ee^{\ii k R_1 -\ii \pi/4}.
\end{equation*}
We can interpret the terms on the right-hand side of~\eqref{eqn:SmallNfrac} in the following way: the first $k^2$ corresponds to the incident wave, the second $4 \nfrac {} \ii \ensem{f_\circ}(0)$ is the contribution from the incident wave scattered once from every cylinder (so-called ``single scattering''), and the last $4 \nfrac {}^2 \ii \ensem{f_{\circ \circ}}(0)$ is the contribution of this scattered wave being re-scattered by every cylinder (so-called ``multiple scattering'').

We can further specialise the wavenumber~\eqref{eqn:SmallNfrac} by considering   wavelengths $2 \pi/k$   larger than the largest cylinder radius, or more precisely
\begin{equation}
k a_* := \max_{\s_1}  \left\{k a_{11}p^2(\s_1) \right\} \ll 1,
\label{cond:ka*}
\end{equation}
which leads to
\begin{equation}
   k_\eff^2 \sim k^2 - 4 \nfrac {} \ii \ensem{f_\circ}(0) + \frac{ 8 \nfrac {}^2}{\pi k^2} \int_0^\pi \cot(\theta/2)  \frac{d}{d \theta}[\langle f_\circ \rangle (\theta)]^2 d\theta + \mathcal O (k^4 a_*^4 \log (k a_*)),
  \label{eqn:EffectiveLowFrequency}
\end{equation}
where the integral converges because $\ensem{f_\circ}'(0) =0$. This expression and the derivation is analogous to that given in \cite[Eq. (86)]{linton_multiple_2005} for a single species. Although the sums in \eqref{eqn:SmallNfrac}   converge quickly for small $k a_*$, the form~\eqref{eqn:EffectiveLowFrequency} is convenient as it is written in terms of the far-field scattering pattern $\ensem{f_\circ}$.
%
%
%

An alternative approach \cite{parnell_multiple_2010}, that is very useful in the context of low frequency propagation is to take the quasi-static limit of the system~\eqref{eqn:AmT}. For small $ka$ the monopole and dipole scattering coefficients are both $O((ka)^2)$, which are the only contributions to the effective bulk modulus and density respectively. Following the approach in \cite{parnell_multiple_2010} it is straightforward to show that for the $N$-species case, where the $n$th species has volume fraction $\phi_n$, bulk modulus $K_n$ and density $\rho_n$, the effective bulk modulus $K_*$ and density $\rho_*$ take the form
\begin{align}
K_*^{-1} &= K^{-1}(\tcr{1 -}\phi)+\sum_{n=1}^N K_n^{-1}\phi_n, &
\rho_* &= \rho\left(\frac{1+\sum_{n=1}^N D_n\phi_n}{1-\sum_{n=1}^ND_n\phi_n}\right),
 \label{eq:lowfreq}
\end{align}
where $\phi = \sum_n \phi_n$ and  $D_n=(\rho_n-\rho)/(\rho_n+\rho)$.

Next we explore how the expression \eqref{eqn:SmallNfrac} compares with other approaches by evaluating it numerically. In Section~\eqref{eqn:AverageCoefficient} we develop analytical expressions for the average scattering coefficient $\A n$ and expressions for the reflection coefficient from the inhomogeneous halfspace.

\section{Two species of cylinders }
\label{sec:TwoTypes}

\begin{figure}[ht]
  \centering
  \begin{subfigure}[h]{0.32\textwidth}		
    \includegraphics[width=\linewidth]{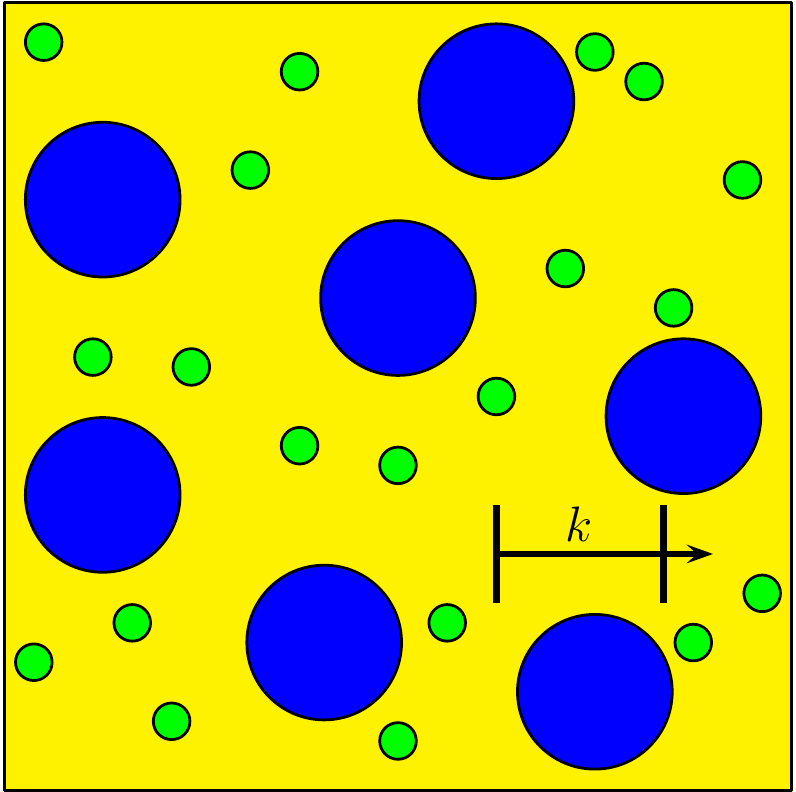}
         \caption{}
        \label{fig:twospeca}
    \end{subfigure}
    \begin{subfigure}[h]{0.32\textwidth}
      \includegraphics[width=\linewidth]{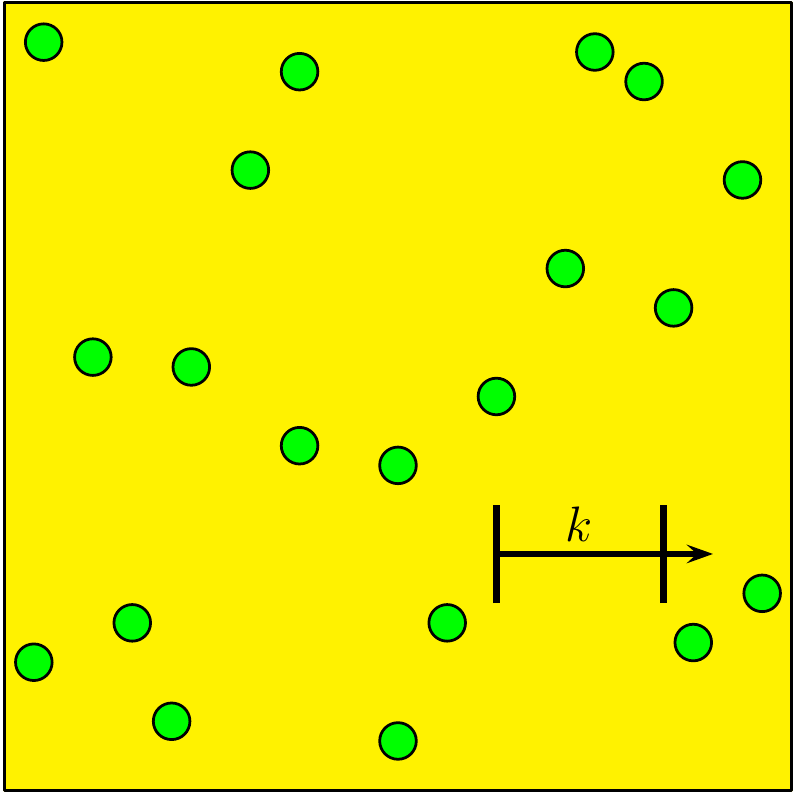}
         \caption{}
        \label{fig:twospecb}
    \end{subfigure}
    \begin{subfigure}[h]{0.32\textwidth}
      \includegraphics[width=\linewidth]{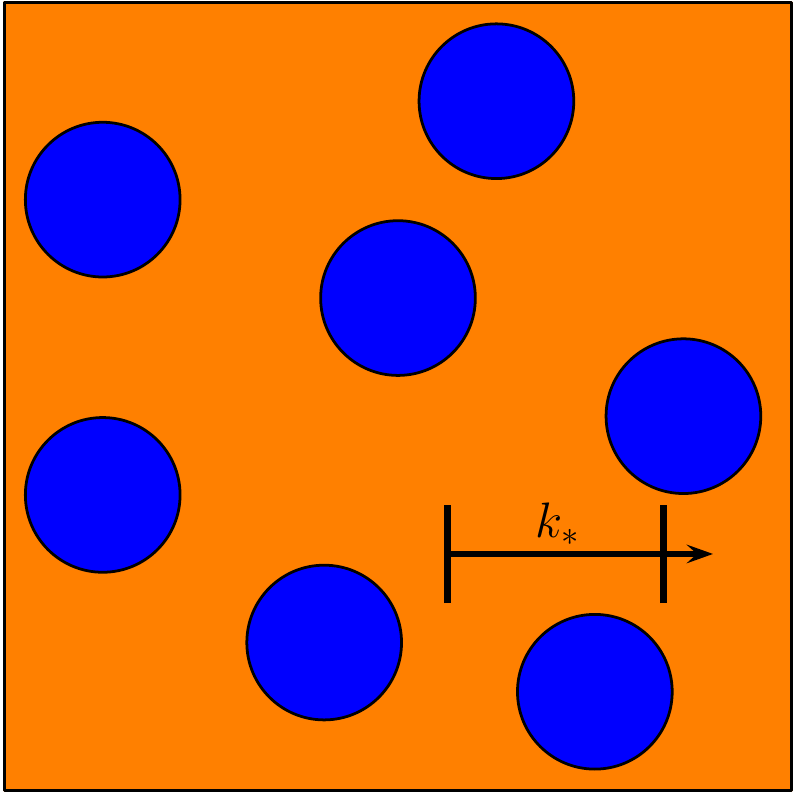}
         \caption{}
        \label{fig:twospecc}
    \end{subfigure}
  \caption{(\subref{fig:twospeca}) Two-species material comprising large (blue) and small (green) inclusions in a background material (yellow) with incident wavenumber $k$, (\subref{fig:twospecb}) one-species material comprising the small inclusions alone, and (\subref{fig:twospecc}) one-species material with large cylinders alone in a background with incident wavenumber $k_\eff$, which is the effective wavenumber of (\subref{fig:twospecb}). }
  \label{fig:twospecall}
\end{figure}

In this section, we analytically compare two approaches to calculating the effective wavenumber of a multi-species mateiral. The first self-consistent type method homogenises the small cylinder distribution and then determines effective properties for a large cylinder distribution embedded in the homogenised background, as shown in Figure \ref{fig:twospecall}. The second determines the multi-species result using the approach outlined in previous sections.


\vspace{-0.8cm} 

\subsection{One small and one large species}
\label{sec:SmallLarge}

We begin by assuming that  there are only two species, $\cs$ and $\cl$, that have constant wave speeds $c_\cs$ and $c_\cl$, densities $\rho_\cs$ and $\rho_\cl$, and number fractions $\nfrac \cs$ and $\nfrac \cl$, respectively. We assume that both types of cylinders have low volume fractions $\phi_\cs = \pi a_\cs^2\nfrac \cs$ and $\phi_\cl = \pi a_\cl^2\nfrac \cl$ and are proportional to one another $\phi_\cs \propto \phi_\cl$, so we will discard $\mathcal O (\phi^3)$ terms, where $\phi = \phi_\cs + \phi_\cl$ denotes the total volume filling fraction. Note that it is more precise to assume small $\phi$, rather than a small number density, since $\phi$ is non-dimensional.

First, the effective wavenumber $k_{\eff \cs}$ of a material at long wavelengths with only a single species of $\cs$-cylinders is obtained by simplifying~\eqref{eqn:EffectiveLowFrequency}, where the far-field pattern~\eqref{eqns:FarFields} is therefore just the $\cs$ cylinder species, i.e.\ there is no integral over $\s_1$ and $p(\s_1) =1$.  Assuming $c_\cs\not =0$ and $\rho_\cs\not =0$, we use  \cite[Eq. (24)]{martin_estimating_2010} for small cylinder radius, which in our notation (recall $\scatZ^n(\s_\ell) := \scatZ_\ell^n$ where $\scatZ_\ell^n$ is given in \eqref{eqn:Zm})
\begin{align}
  &\scatZ^0(\s_\cs) = \ii \pi \frac{a_\cs^2k^2}{4} P  +\mathcal O(a_\cs^4),
  \quad  \scatZ^1(\s_\cs) = \scatZ^{-1}(\s_\cs) = \ii \pi \frac{a_\cs^2 k^2}{4} Q  +\mathcal O(a_\cs^4),
  \label{eqn:zetas_cs}
\\
  &\text{where} \qquad P = 1 - \frac{k^2_\cs \rho}{k^2 \rho_\cs}, \quad   Q= \frac{\rho - \rho_\cs}{\rho + \rho_\cs}.
\end{align}
Substituting the above into the simplified  \eqref{eqn:EffectiveLowFrequency} leads to
\begin{align}
  & \frac{k_{\eff \cs}}{k} = 1  -  \frac{\phi_\cs}{2} (P + 2Q) -\frac{\phi_\cs^2}{8}( 2 P^2 - (P + 2Q)^2)+ \mathcal O(a_\cs^2) + \mathcal O(\phi^3),
\label{eqn:kTcs}
\end{align}
after taking a Taylor series for small $\nfrac {}$ for the square root. We also calculate the effective density \cite[Eq. (1)]{martin_estimating_2010} or refer to \eqref{eq:lowfreq} with $N=1$, given by
\begin{equation}
  \frac{\rho_{\eff \cs}}{\rho} = \frac{\rho + \rho_\cs - \phi_\cs (\rho - \rho_\cs)}{\rho + \rho_\cs + \phi_\cs ( \rho - \rho_\cs)} = 1  -2 \phi_\cs  Q  + \mathcal O(a_\cs^2) + \mathcal O(\phi^3),
\label{eqn:rho_eff}
\end{equation}
which is appropriate for the approximation~\eqref{eqn:kTcs}, see \cite{martin_estimating_2010} for more details.

%

 Next, we determine the effective wavenumber for  large scatterers embedded in a background described by $k_{\eff \cs}$ and $\rho_{\eff \cs}$. For this   step, we
 introduce the notation $f_\circ(0,\s_1) = f_\circ(0,\s_1,\rho,k)$, which expresses the problem in terms of density and wavenumber in place of density and wave speed. Consequently, from \eqref{eqn:rho_eff} we have
\begin{align}
  & f_\circ(0,\s_\cl,\rho_{\eff \cs},k_{\eff \cs}) = f_{\circ\cl}(0)
   -  \phi_\cs \delta f_{\cl \cs} +
 \mathcal O (a_\cs^2) + \mathcal O (\phi^2)
\end{align}
with
\begin{align}
& \delta f_{\cl \cs} := 2 \rho Q \partial_\rho f_\circ(0,\s_\cl,\rho,k) + \frac{k}{2}(P +2  Q)\partial_k f_\circ(0,\s_\cl,\rho,k),
\end{align}
where we set $f_{\circ\cl}(0) := f_\circ(0,\s_\cl,\rho,k)$.

To calculate the wavenumber $k_{*\cl \cs}$ for the $\cl$-cylinders in a material with a wavenumber $k_{\eff \cs}$, we use the formula~\eqref{eqn:SmallNfrac} with $k$ replaced by $k_{\eff \cs}$, $\ensem{f_\circ}$ replaced with $f_\circ(0,\s_\cl,\rho_\cs^*,k_{\eff \cs})$ above and keeping only the integrands, that is, removing the integrals over the multi-species $\s_1$ and $\s_2$, to arrive at
\begin{align}
   k_{\eff \cl \cs}^2 &= k^2_{\eff \cs}
  + 4 \ii \frac{\phi_\cl^2}{\pi a_\cl^4}  \sum_{n,p=-\infty}^\infty a_{\cl \cl}^2 d_{p-n}(k_{\eff\cs} a_{\cl \cl}) \scatZ^n (\s_\cl, \rho_{\eff \cs}, k_{\eff\cs})\scatZ^p (\s_\cl, \rho_{\eff \cs}, k_{T\cs})
  \notag \\
   & \qquad -  \frac{4\ii \phi_\cl}{\pi a_\cl^2}  f_\circ(0,\s_\cl,\rho_{\eff \cs},k_{\eff \cs}) + \mathcal O(\phi^3)
   \label{eqn:SmallNfrac1-cl}
\\
  & =  k^2_{\eff \cs}
  + 4 \ii \frac{\phi_\cl^2}{\pi a_\cl^4}  \sum_{n,p=-\infty}^\infty a_{\cl \cl}^2 d_{p-n}(k a_{\cl \cl}) \scatZ^n (\s_\cl, \rho, k)\scatZ^p (\s_\cl, \rho, k)
\notag\\
  & \qquad -  \frac{4\ii \phi_\cl}{\pi a_\cl^2}  (f_{\circ\cl}(0) -  \phi_\cs \delta f_{\cl \cs}) + \mathcal O(a_\cs^2) + \mathcal O (\phi^3),
  \label{eqn:SmallNfrac-cl}
\end{align}
where we used $d\s_j^m =  \scatZ^m (\s_j)\p(\s_j) d\s_j$.
The above is an attempt to calculate the multi-species wavenumber by using only the single-species formula.
However, the term of order $O(\phi_\cl \phi_\cs)$ in the above does not agree with~\eqref{eqn:SmallNfrac}, even in the limit $a_\cs \to 0$, as we show next.

For only two species of cylinders, and assuming the cylinders are uniformly distributed, the probability density function for the scattering properties becomes
\begin{equation}
  p(\s) = \frac{\nfrac \cs}{\nfrac {}}\delta(a - a_\cs)\delta(c - c_\cs)\delta(\rho - \rho_\cs) + \frac{\nfrac \cl}{\nfrac {}}\delta(a - a_\cl)\delta(c - c_\cl)\delta(\rho - \rho_\cl),
  \label{eqn:pTwo}
\end{equation}
 which when substituted into~\eqref{eqn:SmallNfrac} leads to the multi-species result
\begin{multline}
   k_\eff^2 = k^2 - 4  \ii(\nfrac \cs f_{\circ\cs}(0) + \nfrac \cl f_{\circ\cl}(0))
 +  4  \nfrac \cs^2 a_{\cs \cs}^2 \pi \ii \sum_{n,p=-\infty}^\infty d_{p-n}(k a_{\cs \cs}) \scatZ^p(\s_\cs)\scatZ^n (\s_\cs)
\\
 +  8 \nfrac \cs \nfrac \cl a_{\cs \cl}^2 \pi \ii \sum_{n,p=-\infty}^\infty   d_{p-n}(k a_{\cs \cl}) \scatZ^p(\s_\cs)\scatZ^n (\s_\cl)
\\
+ 4 \nfrac \cl^2  a_{\cl \cl}^2 \pi \ii \sum_{n,p=-\infty}^\infty d_{p-n}(k a_{\cl \cl}) \scatZ^p (\s_\cl)\scatZ^n (\s_\cl) + \mathcal O(\phi^3),
  \label{eqn:SmallNfracTwo}
\end{multline}
where we used
$a_{\cl \cs} = a_{\cs \cl}$. Assuming that $a_\cs \ll 1$ and using $a_{\cl \cs} = b_\cs + b_\cl$, with $b_\cs \geq a_\cs$ and $b_\cl \geq a_\cl$,
we expand $d_m$~\eqref{eqn:d} as
\begin{align}
d_m(k a_{\cs \cl}) = d_m(k b_\cl) + \frac{2 b_\cs}{b_\cl}\left [ J_m(k b_\cl)H_m(k b_\cl) - d_m(k b_\cl)\right ] + \mathcal O(a_\cs^2),
\end{align}
where we use $b_\cs \propto a_\cs$. Substituting the above, \eqref{eqn:zetas_cs}, \eqref{eqn:kTcs} and~\eqref{eqn:SmallNfrac-cl}, into~\eqref{eqn:SmallNfracTwo} we obtain
\begin{equation}
   k_\eff^2 = k^2_{*\cl\cs} +
    \phi_\cl  \phi_\cs \left [ -\frac{4\ii}{\pi a_\cl^2} \delta f_{\cl \cs} + H_0+ \frac{a_\cs}{a_\cl}  H_1 \right ]
     + \mathcal O(a_\cs^2) +\mathcal O (\phi^3),
  \label{eqn:SmallNfracTwoFinal}
\end{equation}
%
where,
\begin{align}
  & G_0 =  \frac{8 \ii}{\pi}\frac{b_\cl^2}{ a_\cl^2} \sum_{n=-\infty}^\infty\sum_{p=-1}^1  d_{p-n}(k b_\cl)  \frac{\scatZ^p(\s_\cs)}{a_\cs^2}\scatZ^n (\s_\cl),
\\
  & G_1 =  \frac{16 \ii}{\pi}\frac{b_\cl b_\cs}{ a_\cl a_\cs}  \sum_{n=-\infty}^\infty\sum_{p=-1}^1  J_{p-n}(k b_\cl)H_{p-n}(k b_\cl) \frac{\scatZ^p(\s_\cs)}{a_\cs^2}\scatZ^n (\s_\cl).
\end{align}
Note that $\scatZ^p(\s_\cs)/a_\cs^2$ converges when $a_\cs \to 0$, see~\eqref{eqn:zetas_cs}.

The terms in the brackets in~\eqref{eqn:SmallNfracTwoFinal} account for the interaction between the two types of cylinders, which is where the  wavenumbers $k_{\eff\cl\cs}$ \eqref{eqn:SmallNfrac-cl} and $k_{\eff}$ \eqref{eqn:SmallNfracTwoFinal} differ. The leading-order error is nonvanishing even as the radius of the small species vanishes, and is given by
\[
\lim_{a_\cs \to 0} \left\{ k_{\eff} - k_{\eff\cl\cs} \right\} \approx \left (G_0  - \frac{4 \ii }{\pi a_\cl^2} \delta f_{\cl\cs} \right)\phi_\cs \phi_\cl.
\]
A numerical investigation of this limit is conducted in Section \ref{sec:Examples}. The physical meaning of these two terms are quite different: $\delta f_{\cl\cs}$ is the change in the far-field scattering pattern of the $\cl$-cylinders due to changing the background wavenumber from $k$ to $k_{*\cs}$, while $G_0$ accounts for the multiple-scattering between the $\cl$ and $\cs$-cylinders, which   becomes significant when both $\phi_{\cs}$
 are $ \phi_{\cl}$ are large. Ultimately, this means  that $k_{*} \approx k_{*\cl\cs}$ if either the $\cs$ or $\cl$-cylinders are very dilute.

For comparison,  we also
give the  2D version of
~\cite[Eq. (23)]{challis_ultrasound_2005} given by
\begin{multline}
   k_{\eff C}^2 = k^2 - 4  \ii \nfrac \cs f_\cs(0) - 4  \ii \nfrac \cl f_\cl(0)
  + \frac{8 \nfrac \cs^2}{\pi k^2} \int_0^\pi \cot(\theta/2) \frac{d}{d \theta}\left[ f_\cs(\theta) \right]^2 d \theta
\\
   + \frac{8 \nfrac \cl^2}{\pi k^2} \int_0^\pi \cot(\theta/2) \frac{d}{d \theta}\left[ f_\cl(\theta) \right]^2 d \theta,
  \label{eqn:NfracTwoChallis}
\end{multline}
and   another commonly used approximation~\cite{ma_application_1984}:
\begin{equation}
   k_{\eff0}^2 = k^2 - 4  \ii \nfrac \cs f_\cs(0) - 4  \ii \nfrac \cl f_\cl(0),
  \label{eqn:oneNfracTwo}
\end{equation}
describing the effective wavenumber for two species. However, the expression~\eqref{eqn:NfracTwoChallis} is missing the interaction between the $\nfrac \cs$ and $\nfrac \cl$ species (the term $\mathcal O(\nfrac \cs \nfrac \cl)$) and is only valid for low frequency. Additionally, the estimate~\eqref{eqn:oneNfracTwo} ignores terms of the order $\mathcal O (\phi^2)$.


 \section{Numerical Examples}
 \label{sec:Examples}

In this section, we consider a selection of numerical examples to demonstrate the efficacy of \eqref{eqn:SmallNfracTwo} and other expressions.  For $k_{\eff\cl \cs}$ we use the exact formula~\eqref{eqn:SmallNfrac} for one-species and then equation~\eqref{eqn:SmallNfrac1-cl}. This way, $k_{\eff\cl \cs}$ is valid for $\cs$-cylinders \edit{with} approximately zero density such as air. In the graphs that follow we use
\begin{equation}
  \text{sound speed } = \frac{\omega}{\mathrm{Re} \, k_\eff} \quad \text{and} \quad
  \text{attenuation } = \mathrm{Im} \, k_\eff,
\end{equation}
where $k_\eff$ will be replaced with $k_{\eff\cl \cs}$ and $k_{\eff0}$ depending on the context.

For reference, we provide Julia~\cite{bezanson_julia:_2017} code to calculate the effective wavenumbers.

\subsection{2D Emulsion}

\begin{table}[h]
\centering
\begin{tabular}{l|| c | c | c| c|}
   & density (kg/m${}^3$) & speed (m/s) & radius ($\mu$m) & volume \% \\ [0.5ex]
\hline\hline
Distilled Water & $\rho = 998$ & $c = 1496$ & -- & 84 \% \\
\hline
Hexadecane & $\rho_\cl = 773$ & $c_\cl =1338$ & $a_\cl =  250$ & $\phi_\cl = 11\%$\\
\hline
Glycerol & $\rho_\cs = 1260$ & $c_\cs = 1904$ & $a_\cs = 25$ & $\phi_\cs = 11\%$
\end{tabular}
\caption{Material properties used to approximate an emulsion.}
\label{tab:fluid}
\end{table}

\begin{figure}[]
  \centering
  \begin{subfigure}[a]{\textwidth}
    \includegraphics[width=0.95\linewidth]{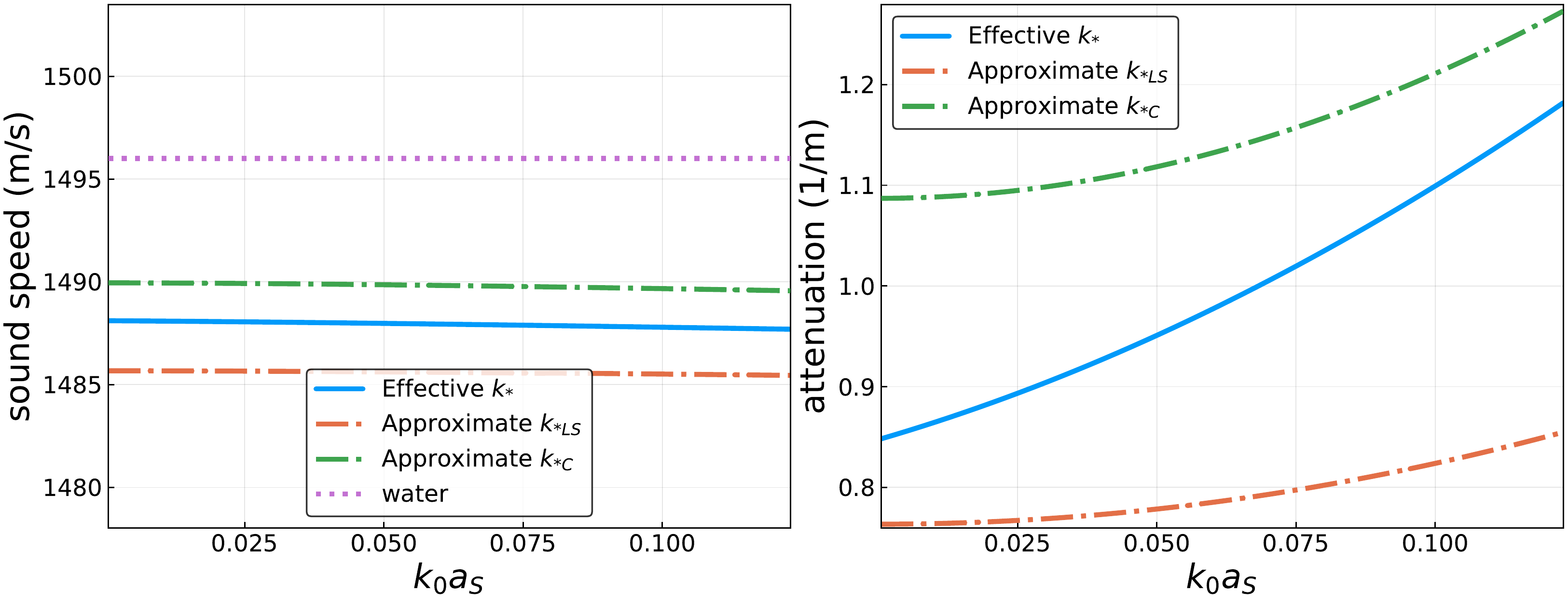}
        \caption{fixed wavenumber $k = k_0$ while changing the radius $a_\cs$.}
        \label{fig:fluid_small}
    \end{subfigure}
    \begin{subfigure}[b]{\textwidth}
      \includegraphics[width=0.95\linewidth]{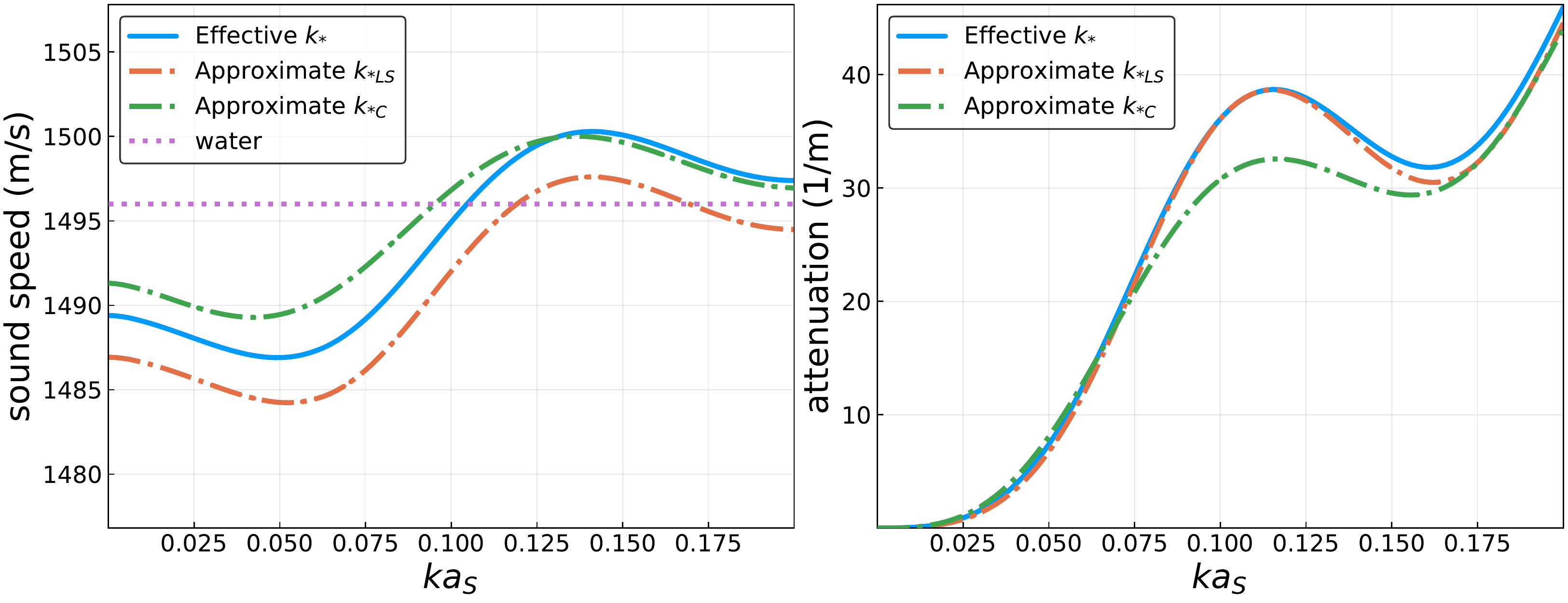}
        \caption{fixed radius $a_\cs$ while changing the wavenumber $k$}
        \label{fig:fluid_compare}
    \end{subfigure}
    \begin{subfigure}[c]{\textwidth}
      \includegraphics[width=0.95\linewidth]{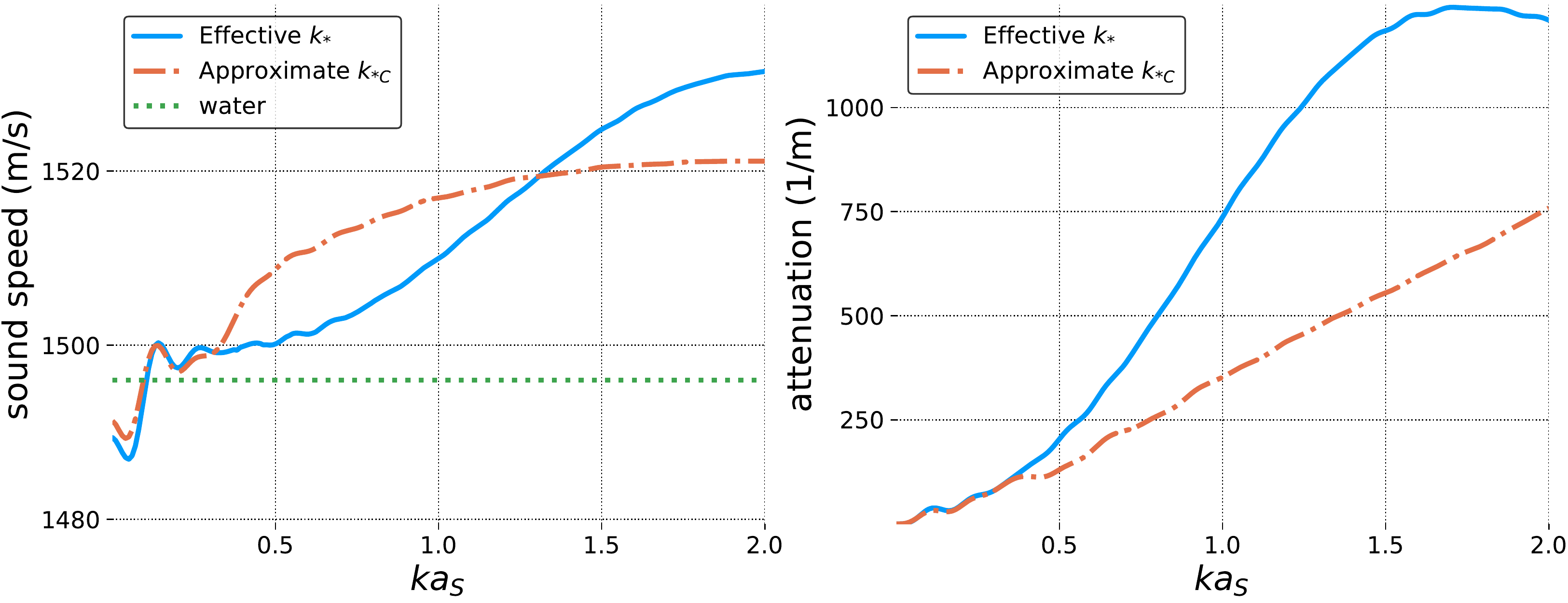}
        \caption{fixed radius $a_\cs$ while changing the wavenumber $k$}
        \label{fig:fluid_large-w}
    \end{subfigure}
  \caption{Comparison of sound speeds and attenuation using wavenumbers $k_{\eff C}$~\eqref{eqn:NfracTwoChallis}, $k_{\eff \cl \cs}$~\eqref{eqn:SmallNfrac1-cl}, and   $k_{\eff}$~\eqref{eqn:SmallNfracTwo} for the water and
  oil emulsion from Table~\ref{tab:fluid}. Code to generate figure: \url{https://github.com/arturgower/EffectiveWaves.jl/tree/master/examples/emulsion}.
}
  \label{fig:fluid}
\end{figure}

Here we consider an emulsion composed of Hexadecane (oil) and Glycerol in water~\cite{magdassi_formation_1986}, see Table~\ref{tab:fluid}, where  the   glycerol forms very small inclusions.
The graphs of Figure~\ref{fig:fluid_small} show how $k_{\eff}$, $k_{\eff \cl \cs}$ and $k_{\eff C}$ differ when varying only $a_\cs$ the radius of the glycerol inclusions, for a fixed angular frequency $\omega = c/k_0 \approx 3 \times 10^6 \, \mathrm{Hz}$. We observe that the difference between $k_\eff$ and $k_{\eff\cl\cs}$ persists even as
 $a_\cs \to 0$, as expected according to~\eqref{eqn:SmallNfracTwoFinal}. Meaning that, no matter how small the $\cs$-cylinders  become, the larger cylinders $\cl$ do not perceive the $\cs$-cylinders as a homogeneous material, in the naive way described in Section~\ref{sec:TwoTypes}\ref{sec:SmallLarge}.

In \cite[Fig. 21]{challis_ultrasound_2005}, they observed that experimentally measured \edit{wave} speeds were shifted in comparison to the $k_{\eff C}$ predictions, even for low-frequency.   We can see this same discrepancy in Figure~\ref{fig:fluid_compare}, where the angular frequency is varied between $1\, \mathrm{KHz} < \omega < 12 \, \mathrm{MHz}$ while the radius $a_\cs = 25\, \mu$m is fixed.
This discrepancy is   due to the terms of order $O(\nfrac \cl \nfrac \cs)$   which are missing from $k_{\eff C}$~\eqref{eqn:NfracTwoChallis} and $k_{\eff \cl \cs}$ \eqref{eqn:SmallNfracTwo}. Although  all three wavenumbers are     similar in Figure~\ref{fig:fluid_compare}.  The same is not true when we increase the frequency.

In Figure~\ref{fig:fluid_large-w} we show how $k_{\eff C}$, valid only for low-frequency, strays from the more accurate $k_\eff$ as the frequency increases\footnote{In this case, we did not exactly use the 2D version of equation (23) from~\cite{challis_ultrasound_2005}, but instead used a more accurate version where we summed enough terms for the far-field patterns to converge.}, where we did not include $k_{\eff \cl \cs}$ as it is only valid for low frequency. There we can see that
$k_{\eff C}$ performs well up to about $k a_\cs = 0.3$, at which point $k a_\cl = 3.0$.

\begin{figure}[h]
  \centering
  \includegraphics[width=0.98\linewidth]{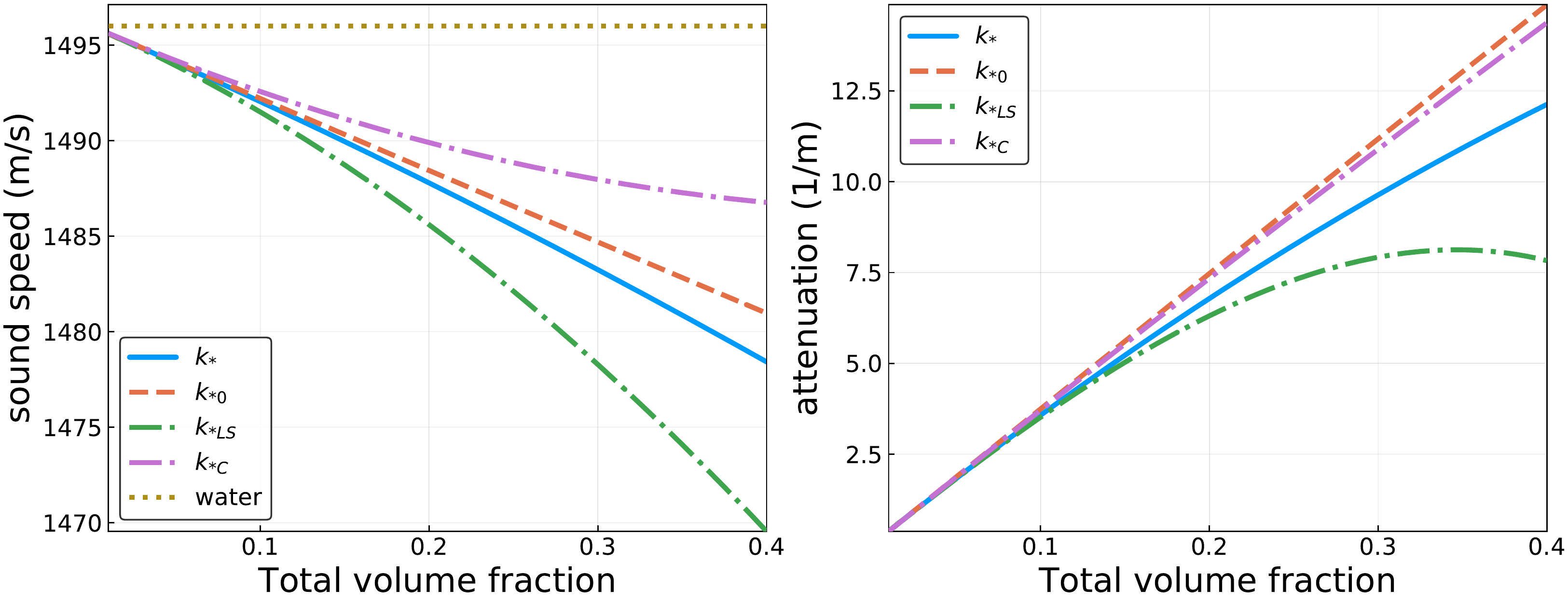}
  \caption{Comparison of sound speeds and attenuation calculated from   the   effective wavenumbers $k_{\eff C}$~\eqref{eqn:NfracTwoChallis}, $k_{\eff0}$~\eqref{eqn:oneNfracTwo},
  $k_{\eff \cl \cs}$~\eqref{eqn:SmallNfrac1-cl} and the more accurate $k_\eff$~\eqref{eqn:SmallNfracTwo}, as the total volume fraction of the inclusions increases (for the emulsion shown in Table~\ref{tab:fluid} with $k a_\cs = 0.5$). Code: \url{https://github.com/arturgower/EffectiveWaves.jl/tree/master/examples/emulsion}.
  }
  \label{fig:fluid_volfrac}
\end{figure}

All the approximations $k_{\eff0}$~\eqref{eqn:oneNfracTwo},
$k_{\eff \cl \cs}$~\eqref{eqn:SmallNfracTwo} and $k_{\eff C}$~\eqref{eqn:NfracTwoChallis} are missing second order terms in the number density. In Figure~\ref{fig:fluid_volfrac} we see the effect of these missing terms by varying the volume fraction while fixing $\omega = 3 \times 10^6$
, or equivalently $k a_\cs = 0.5$. In the limit of low volume fraction, all three effective wavenumbers agree, as expected. For the largest volume fraction
$40\%$, the expected error\footnote{If we disregard the error due to the low-frequency assumptions, we can estimate the expected errors $\phi^2 = 0.4^2 =16\%$ for $k_{\eff0}$ and $2\phi_\cl \phi_\cs = 2*0.2^2 = 8\%$ for both $k_{\eff \cl \cs}$ and $k_{\eff C}$} \hspace{-0.2cm}
of $k_{\eff}$ is $6\%$.
However, the relative difference between the attenuations of $k_{\eff \cl \cs}$ and $k_{\eff 0}$ and the multi-species attenuation of $k_{\eff}$ reaches $30\%$.

Summarising Figures~\ref{fig:fluid} and \ref{fig:fluid_volfrac}, all the approximations are similar for either low frequency or low volume fraction. This is because the three phases in Table~\ref{tab:fluid} have similar properties. In our next example one of the phases, air, will be very different from the others, which will lead to more dramatic differences.

\subsection{2D Concrete}

When there is a  high contrast in the properties of the inclusions, multiple scattering can have a dramatic effect. To demonstrate this we consider a concrete-like material made from a limestone possessing cylinders of brick and air, given in Table~\ref{tab:concrete}.
\begin{table}[]
\centering
\begin{tabular}{l|| c | c | c| c|}
   & density (kg/m${}^3$) & speed (m/s) & radius (mm) & volume \% \\ [0.5ex]
\hline\hline
Limestone & $\rho = 2460$ & $c = 4855$ & -- & 84 \% \\
\hline
Brick & $\rho_\cl = 1800$ & $c_\cl =3650$ & $a_\cl =  2.0$ & $\phi_\cl = 10\%$\\
\hline
Dry air & $\rho_\cs = 1.293$ & $c_\cs = 331.4$ & $a_\cs = 0.2$ & $\phi_\cs = 6\%$
\end{tabular}
\caption{Material properties for our concrete-like material. Note that we used compacted limestone with very low porosity~\cite{parent_mechanical_2015}.}
\label{tab:concrete}
\end{table}

\begin{figure}
  \centering
  \begin{subfigure}[a]{\textwidth}
    \includegraphics[width=0.95\linewidth]{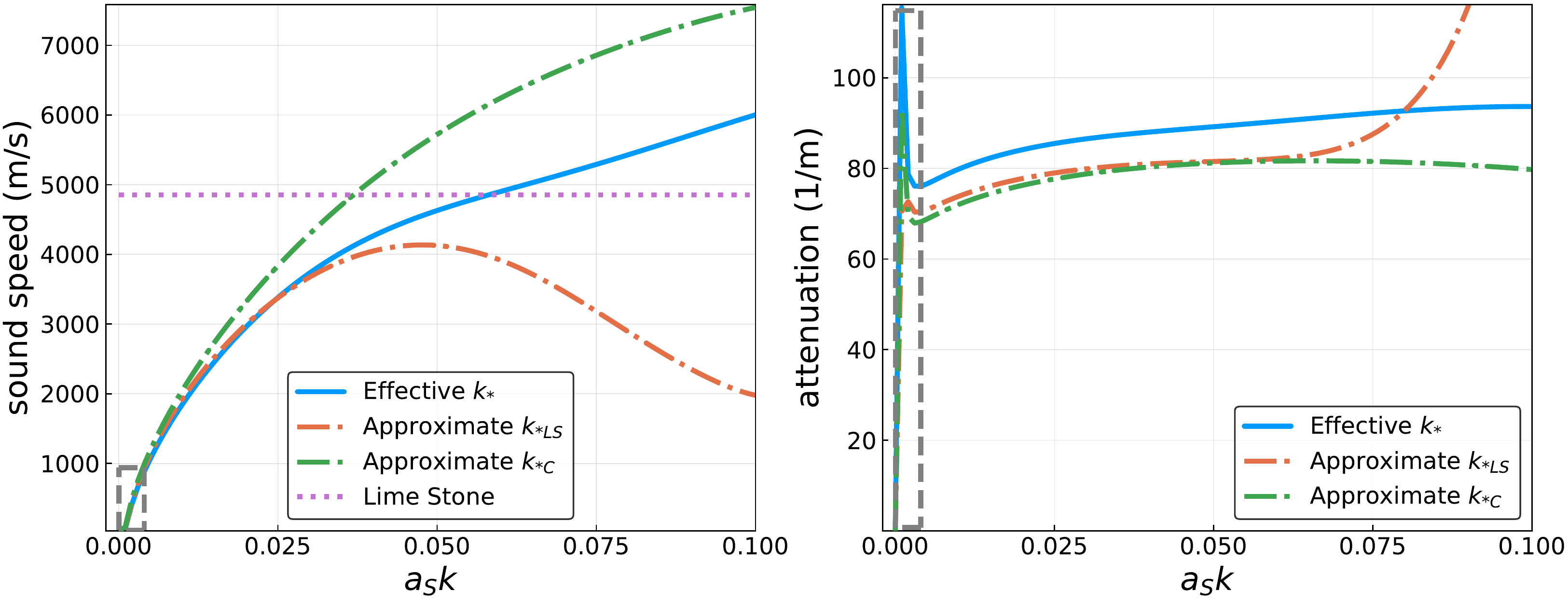}
        \caption{fixed wavenumber $k = k_0$ while changing the radius $a_\cs$.}
        \label{fig:concrete_compare}
    \end{subfigure}
    \begin{subfigure}[b]{\textwidth}
      \includegraphics[width=0.95\linewidth]{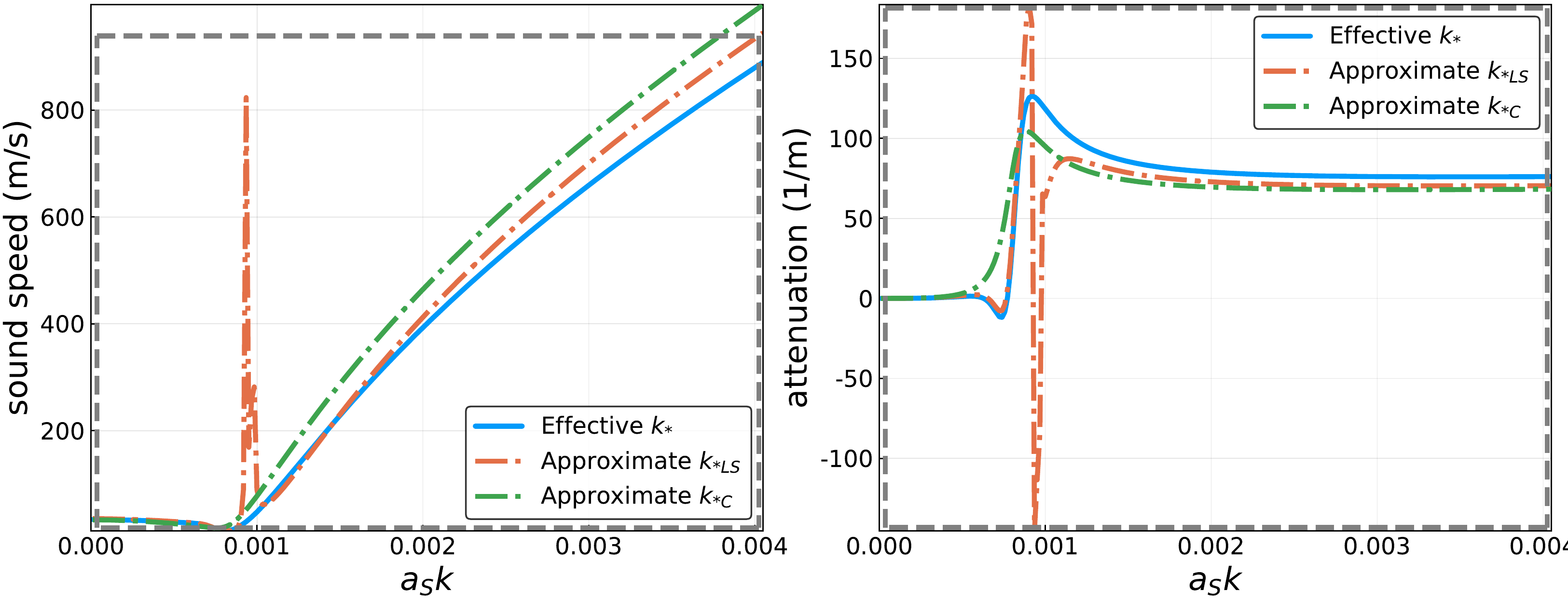}
        \caption{Same as (a) but for smaller $a_\cs k$ }
        \label{fig:concrete_compare_zoom}
    \end{subfigure}
    \begin{subfigure}[c]{\textwidth}
      \includegraphics[width=0.95\linewidth]{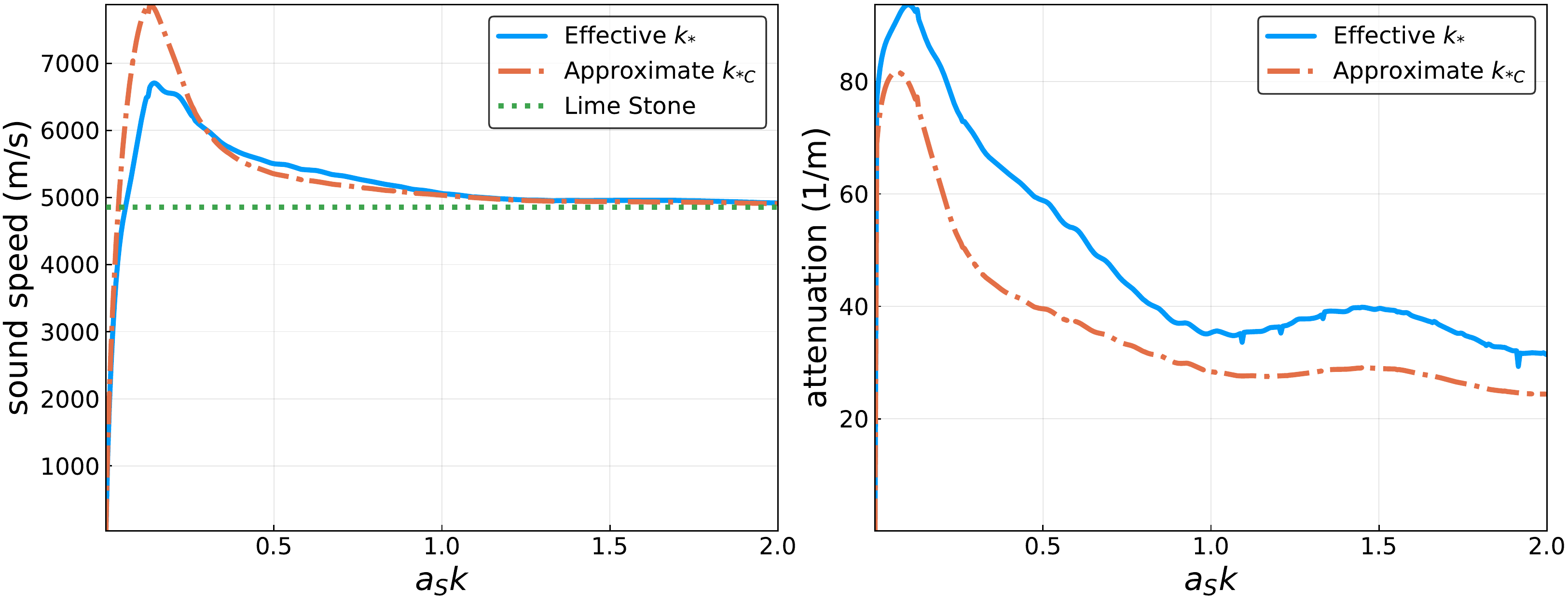}
        \caption{fixed radius $a_\cs$ while changing the wavenumber $k$}
        \label{fig:concrete_large-w}
    \end{subfigure}
  \caption{  Sound speed and attenuation from the approximate wavenumbers $k_{\eff C}$~\eqref{eqn:NfracTwoChallis} and $k_{\eff \cl \cs}$~\eqref{eqn:SmallNfrac1-cl} with the more accurate $k_{\eff}$~\eqref{eqn:SmallNfracTwo} for the concrete-like mixture shown in Table~\ref{tab:concrete}. Code: \url{https://github.com/arturgower/EffectiveWaves.jl/tree/master/examples/concrete}.
  }
  \label{fig:concrete}
\end{figure}

Figure~\ref{fig:concrete} shows that it is  only in the low frequency limit, $k a_\cs < 0.05$, that the wavenumbers $k_{\eff C}$ and $k_{\eff\cl \cs}$ agree with the more exact $k_{\eff}$, which has a maximum expected relative error of only $\phi^3 = 0.16^3 \approx 0.4\%$. And in Figure~\ref{fig:concrete_compare_zoom} the wavenumber $k_{\eff \cl \cs}$
 appears to hit a resonance which should not be present. This, and the dramatic changes in attenuation   at low frequency, are   expected because for an inclusion with low density,   the effective wavenumber diverges for fixed volume fraction when $k$ tends to zero~\cite{martin_estimating_2010}.
 Figure~\ref{fig:concrete_large-w} shows the limitations of $k_{\eff C}$ as the frequency increases. Even though $k_{\eff C}$ is only valid for low frequencies, its results are quite close to $k_{\eff}$, having a relative difference of around $25\%$.

\begin{figure}
  \centering
  \includegraphics[width=0.98\linewidth]{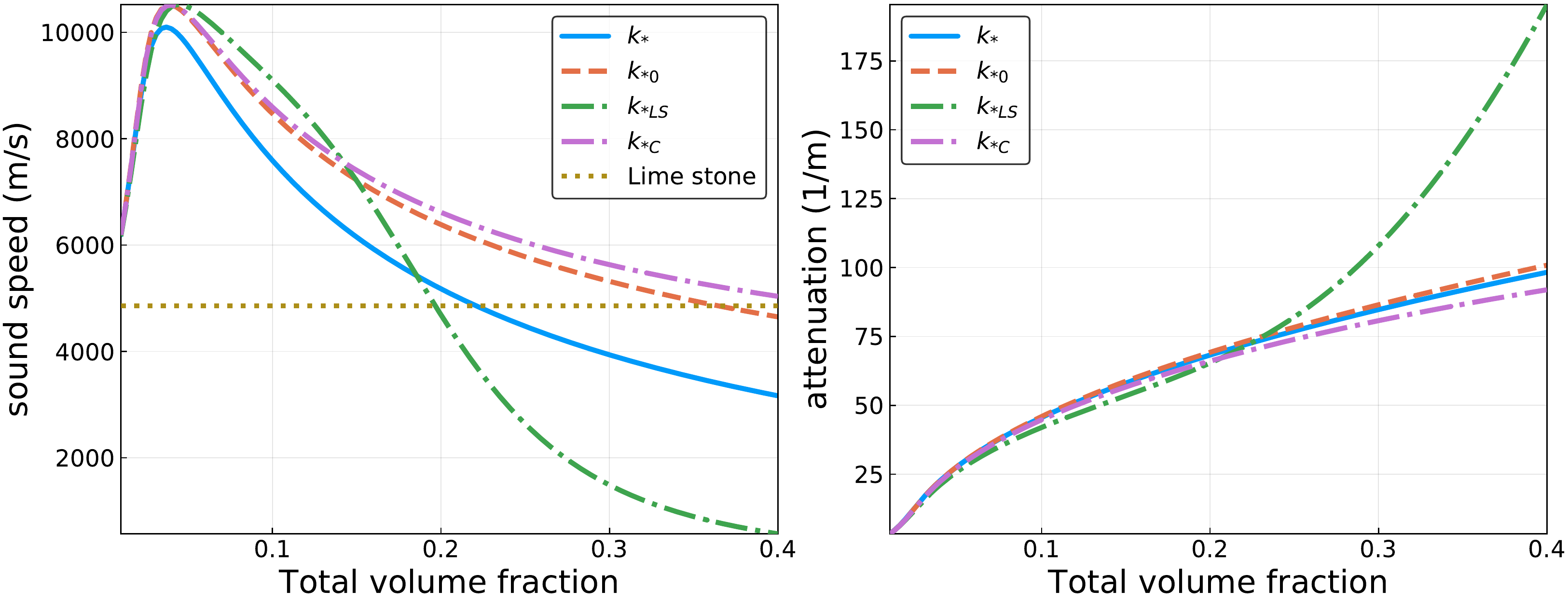}
  \caption{Sound speed and attenuation from the three effective wavenumbers $k_{\eff C}$~\eqref{eqn:NfracTwoChallis}, $k_{\eff0}$~\eqref{eqn:oneNfracTwo} and   $k_{\eff \cl \cs}$~\eqref{eqn:SmallNfrac1-cl}, with the more accurate $k_{\eff}$~\eqref{eqn:SmallNfracTwo}, against the total volume fraction of the inclusions   for the concrete shown in Table~\ref{tab:concrete}.
}
  \label{fig:concrete_volfrac}
\end{figure}

Again as expected, all the wavenumbers converge as the volume fraction decreases, see Figure~\ref{fig:concrete_volfrac}, yet the differences in the \edit{wave} speed are significant, reaching $100\%$ in this example, when the total volume fraction $\phi = 40\%$.

\section{The average field and reflection}
\label{eqn:AverageCoefficient}

In this section we determine the reflected field from a half-space, which is achieved by deducing the averaged scattering coefficient $\A n$, for low number density shown in \eqref{eqns:nfracExpansion}${}_2$.

In order to calculate   $\A n$ we first use~\eqref{eqn:AmInc} and~\eqref{eqn:SmallNfrac}, except here we \edit{can} deduce that $K_{\eff 0} = k^2$. As $\theta_{\eff}$ appears in \eqref{eqn:AmInc} we expand
\begin{equation}
\theta_\eff= \theta_{\eff 0} + \theta_{\eff 1} \nfrac {} + \theta_{\eff 2} \nfrac {}^2 + \mathcal O(\nfrac {}^2),
\label{eqn:thetaExpansion}
\end{equation}
which combined with Snell's equation~\eqref{eqn:Snells}
 and the number density expansions \eqref{eqns:nfracExpansion} gives for the first two orders:
\begin{equation}
  K_{\eff 0} \sin (\theta_{\eff 0})^2 = k^2 \sin (\theta_\inc)^2,\quad
  k^2 \theta_{\eff 1} \cos(\theta_{\eff 0}) = 2 \ii \ensem{f_\circ}(0) \frac{\sin(\theta_{\eff 0})^3}{\sin(\theta_{\inc})^2}.
\label{eqn:SnellLow}
\end{equation}

For $\bar x$, which appears in equation \eqref{eqn:AmInc}, we assume that, as $\nfrac {} \to 0$,  $\bar x$ is a fixed width large enough for the effective wave ansatz~\eqref{eqn:AnsatzA} to hold, meaning that
\begin{equation}
  \bar x =  \mathcal O(1), \quad
  \int_0^{\bar x} \A n(x_2,\s)\ee^{-\ii \alpha x_2} d x_2 = \mathcal O(1).
\end{equation}
Using the above in \eqref{eqn:AmInc} we conclude that $K_{\eff 0} = k^2$, to ensure that $\nfrac {}^{-1}$ appears on the left-hand side. Subsequently, using $K_{\eff 0} = k^2$ in~\eqref{eqn:SnellLow} leads to
\begin{align}
  \theta_{\eff 0} &= \theta_\inc, \quad \theta_{\eff 1} = \frac{2\ii \ensem{f_\circ}(0)}{k^2}\tan \theta_\inc,
  \\
\label{eqns:thetaT}
\theta_{\eff 2} &=  \theta_{\eff 1}^2\left[\frac{\cos \theta_\inc}{\sin\theta_\inc} +\frac{1}{\sin (2 \theta_\inc)}\right] + 2 \ii \frac{\ensem{f_{\circ\circ}}(0)}{ k^2} \tan \theta_\inc,
\end{align}
where $\theta_{\eff 2}$ is given   for completeness. We use the above to expand:
\begin{equation}
  \frac{\ee^{\ii (\alpha_\eff-\alpha){\bar x}}}{\alpha_\eff- \alpha} = i \bar x + \left [\frac{1}{\nfrac {} } - \frac{ \ensem{f_{\circ \circ}}(0)}{ \ensem{f_\circ}(0)}  \right ]\frac{\ii k}{2 \ensem{f_\circ}(0)} \cos \theta_\inc    +\frac{1}{2 k} \sec \theta_\inc + \mathcal O(\nfrac {})
\label{eqn:pole_expand}
\end{equation}
then we substitute the leading order term in the above into \eqref{eqn:AmInc} leading to
:  $
\ensem{f_\circ}(0) = \sum_{n=-\infty}^\infty  \int_\regS \A n_{\eff 0} d\s_2^n
$. However, from \eqref{eqn:K1} we found that $\A n_{\eff 0}$ is independent of $n$ and $\s_2$, therefore $\A n_{\eff 0} = -1$.
This means that $\A n_\eff$ tends, in the limit $\nfrac {} \to 0$, to the scattering coefficient of one lone cylinder:
\begin{equation}
 \A m(\mathbf x_1 ,\s_1) \; \to \; \ii^m \A m_{\eff 0}(\s_1) \ee^{-\ii m \theta_{{\eff 0}}} \ee^{\ii \mathbf x_1 \cdot \sqrt{ \mathbf K_{\eff 0}}} = -\ii^m \ee^{-\ii m \theta_\inc} \ee^{\ii \mathbf x_1 \cdot \mathbf k} = A^m_1,
 \label{eqn:SingleScatterer2}
\end{equation}
where we used $K_{\eff 0} = k^2$ and $\theta_{\eff 0} = \theta_{\inc}$, and the last equation is from \eqref{eqn:SingleScatterer}.

To calculate the next order in $\nfrac {}$ of equation \eqref{eqn:AmInc} we need to calculate $b(\bar x)$. To do so, we assume that $\A n(\mathbf x_2,\s_2) \to A^n_2$ as $\nfrac {} \to 0$ for every $x_2 >0$. That is, in the limit where there are no cylinders, except one fixed at $\mathbf x_2$, the averaged  scattering coefficient $\A n$ tends to the the scattering coefficient of one lone cylinder, even for $0< x_2 < \bar x$. As a result
\[
   \A n(x_2,0,\s_2)  = -\ii^n \ee^{-\ii n \theta_\inc} \ee^{\ii \alpha x_2 } + \mathcal O(\nfrac {}), \quad \text{for} \;\; x_2>0,
\]
which when substituted in $b(\bar x)$ from \eqref{eqn:AmInc}, together with~\eqref{eqns:FarFields}, leads to $b(\bar x) =   \ii \bar x \ensem{f_\circ}(0) + \mathcal O(\nfrac {})$. Substituting this,~\eqref{eqn:pole_expand}, and \eqref{eqns:thetaT} into~\eqref{eqn:AmInc}, and then ignoring second order terms $\mathcal O(\nfrac {}^2)$ we obtain
\begin{multline}
 \frac{ - \ii \ensem{f_{\circ \circ}}(0) k \cos \theta_\inc}{2 \ensem{f_\circ}(0)}
 + \frac{\ii k \cos \theta_\inc}{2 \ensem{f_\circ}(0)}
  \sum_{n=-\infty}^\infty \int_\regS
  \A n_{\eff 1}(\s_2)
  d\s_2^n
 =
    -\frac{\ensem{f_\circ}(0)}{2k\cos \theta_\inc} + \mathcal O (\nfrac {}^2),
    \label{eqn:AT1-1}
\end{multline}
where we also used
$
  \sum_{n=-\infty}^\infty \int_\regS
     n d\s_2^n = 0,
$
which is a result of the property $\scatZ^n_2 = \scatZ^{-n}_2$, see \eqref{eqn:Zm}, implying that $d\s_2^{n} = d\s_2^{-n}$.
In Appendix~\ref{sec:Integrals}\ref{sec:app_LowNumberFraction},
we showed  that the quantity $F_\eff$, given by~\eqref{eqn:independent-FT},
is independent of $n$ and $\s_2$. So if we substitute $\A n_{\eff 1}(\s_2)$ for $F_\eff$, we can then take $F_\eff$ outside the sum and integral in~\eqref{eqn:AT1-1}, and then substitute back $\A n_{\eff 1}(\s_2)$ to arrive at
\begin{equation}
   \A n_{\eff 1}(\s_2) = - \frac{\ii \ensem{f_\circ}(0)}{ k^2 \cos^2 \theta_\inc}
  - \pi \sum_{m=-\infty}^\infty \int_{\regS} a_{12}^2 d_{m-n}(k a_{12}) d \s_1^m + \mathcal O (\nfrac {}^2),
    \label{eqn:AT1}
\end{equation}
where we used~\eqref{eqns:FarFields}${}_2$. The above reduces to the one species case given by \cite[Eq. (27)]{martin_multiple_2011}. With $\A n_{\eff 1}$ and $\A n_{\eff 0}$ we can now calculate reflection from a halfspace.

\subsection{Reflection from a halfspace}
\label{sec:Reflection}

 Here we calculate the reflected wave measured at $(x,y)$, where $x <0$. 
 To achieve this we assume that the boundary layer around $x =0$ has little effect on the reflected wave, that is, we assume most of the scatterers behave as if they are in an infinite medium. This is the same as taking $\bar x =0$, which was also used in~\cite{martin_multiple_2011}, \add{where they showed that this approach matches other homogenisation results in the low-frequency limit. We note however that \cite{felbacq_scattering_1994} discusses the possibility of a boundary layer effect even in the low-frequency limit.}

\begin{figure}[ht!]
  \centering
  \includegraphics[width=0.5\textwidth]{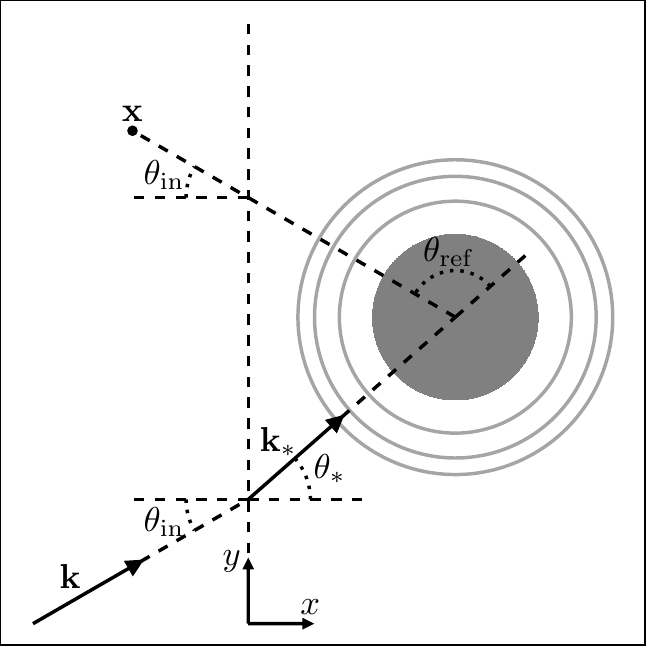}
  \caption{illustrates the far-field reflected angle $\theta_\reflect$, where $\mathbf k = (\alpha,\beta)$ and $\mathbf k_\eff$ is the effective transmitted wavenumber defined in Section~\ref{sec:EffectiveWavenumbers}.
 The wavenumber $\mathbf k_\eff$ results from ensemble averaging all scattered waves originating from $x>0$ (to the right of the dashed vertical line). The reflected field measured at $\mathbf x$ can be understood as the scattering (the grey circles) of the transmitted wave by an effective particle (grey particle).
In the figure, $\theta_\reflect$ equals $\pi - \theta_\eff - \theta_\inc$, but for small number density $\theta_\eff = \theta_\inc + \mathcal O(\nfrac {})$, which is why $\theta_\reflect = \pi - 2\theta_\inc$ appears in \eqref{eqn:R1} and~\eqref{eqn:R2}.
 }
  \label{fig:reflect_farfield}
\end{figure}

Substituting~\eqref{eqn:AverageWaveCond} into the total effective wave~\eqref{eqn:AverageWave}, and using the form~\eqref{eqn:AnsatzA} reveals
\begin{multline}
  \ensem{u(x,y)} = \ee^{\ii \vec k  \cdot \vec x}
  + \nfrac {} \ee^{-\ii m \theta_\eff} \sum_{m=-\infty}^\infty \ii^m   \int_\regS \A m_\eff( \s_1) \int_{0 < x_1< \infty}
\ee^{\ii \beta y_1+\ii \alpha_\eff x_1} \Phi_m(k R_1,\Theta_1) d \vec x_1 d \s_1^m,
\label{eqn:TotalReflection}
\end{multline}
where we used $u_\inc(x,y) = \ee^{\ii \vec k  \cdot \vec x}$, $d \s_1^m = \scatZ_1^m p(\s_1) d \s_1$, $\Phi_m(k R_1,\Theta_1) = H_m^{(1)}(k R_1) \ee^{\ii m \Theta_1}$, substituted \eqref{eqn:pLam1},  used $N =  |\reg| \nfrac {}$ and took the limit $N \to \infty$.
Then using~\eqref{eqn:McircAndline} and \eqref{eqn:Mline} we obtain
\begin{equation}
  \int_{0 < x_1< \infty}
\ee^{\ii \beta y_1+\ii \alpha_\eff x_1} \Phi_m(k R_1,\Theta_1) d \vec x_1=
   \ee^{ -\ii\alpha x + \ii\beta y} \frac{2}{\alpha}\frac{(-\ii)^{-m}\ii}{\alpha + \alpha_\eff}
   \ee^{-\ii m \theta_\inc} ,
\end{equation}
noting that $x_1-x>0$. Using the above in~\eqref{eqn:TotalReflection} we reach
\begin{equation}
  \ensem{u(x,y)}
   =
  \ee^{\ii \vec k  \cdot \vec x}
  + \frac{2\nfrac {}}{\alpha}
   \frac{\ii \ee^{-\ii \alpha x + \ii \beta y} }{\alpha + \alpha_\eff}
       \sum_{m=-\infty}^\infty \ee^{ \ii m \theta_\reflect} \int_\regS
  \A m_\eff(\s_1) d \s_1^m ,
  \label{eqn:TotalReflection2}
\end{equation}
where $\theta_\reflect = \pi - \theta_\eff - \theta_\inc$.
The reflected wave shown by~\eqref{eqn:reflection_results} is calculated by expanding for small $\nfrac {}$, including $\theta_\reflect = \pi  - 2\theta_\inc + \mathcal O(\nfrac {})$, and then substituting the results from Section~\ref{eqn:AverageCoefficient}.
For a single-species this formula reduces to   \cite[Eqs. (41) and (42)]{martin_multiple_2011}. The Figure~\ref{fig:reflect_farfield} gives a pictorial representation of the reflection coefficient in~\eqref{eqn:TotalReflection2}.

\section{Conclusion}
\label{sec:conclusion}
We have deduced the effective wavenumbers~\eqref{eqn:wavenumber_results} and \eqref{eqn:SmallNfracSpheres}, and reflection coefficient~\eqref{eqn:reflection_results},  for a  multi-species material up to moderate number density and over a broad range of frequencies. This will enable experimental researchers to extract more information about the makeup of inhomogeneous media (see the supplementary material for self-contained expressions for the wavenumbers and reflection coefficients in the case of a finite number of species). We also remark that the results may be extended straightforwardly to multiple scattering from cylinders in a number of contexts, including two-dimensional electromagnetism.

Characterisation is not the only application; this theory can also be employed to design novel materials. We have shown that multiple scattering between different species can lead to effective properties that are not exhibited by single species media. That is, using our multi-species formulas it is now possible to choose species so as to design impedance matched, highly dispersive and broad band attenuating materials.

We also saw that the multi-species effective wavenumbers derived in the acoustics literature were accurate for low frequency and low volume fraction. But to go beyond these limitations, our more precise effective wavenumber was needed. We also illustrated that a ``self-consistent'' approach to calculating the effective wavenumber is not even accurate at low frequencies.


Two main issues of our method deserve further investigation: the effects of the boundary layer near the surface of the halfspace, and the quasicrystalline approximation.
To calculate the reflection coefficient up to second order in the number density, we neglected the effects of the boundary layer. It is not clear how to theoretically make progress without these two approximations, nor what errors they introduce. We believe that these issues represent important future work.

\appendix
\renewcommand{\theequation}{\Alph{section}.\arabic{equation}}

\section{Effective wavenumber for multi-species spherical inclusions}
\label{sec:SphericalWavenumber}

In this section, we apply our multi-species theory to the results in \cite{linton_multiple_2006} for spherical inclusions to reach the effective wavenumber~\eqref{eqn:SmallNfracSpheres}. Details are omitted when the results follow by direct analogy. For spheres we define the ensemble-average far-field pattern and multiple-scattering pattern,
\begin{align}
  &\ensem{F_\circ}(\theta) = - \sum_{n=0}^\infty P_n (\cos \theta ) \int_{\regS} (2n+1) d \s^n_1,
\\
  &\ensem{F_{\circ\circ}} = \frac{\ii (4 \pi)^2}{2} \sum_{n=0}^\infty \sum_{p=0}^\infty \sum_q \int_{\regS}\int_{\regS}  \frac{\sqrt{(2n+1)(2p+1)}}{(4 \pi)^{3/2}}  \sqrt{2 q +1}  \mathcal G(n,0;p,0;q) k a_{12} D_q(k a_{12})  d \s_1^n d \s_2^p, 
\label{eqns:SphericalPatterns}
\end{align}
where $q$ takes the values
\[
|n-p|, |n-p|+2, |n-p|+4, \ldots, n+p,
\]
$d\s^n_i =  \scatZs^n(\s_i) p(\s_i) d \s_i$, $D_m(x) = x j_m'(x)( x h_m'(x) + h_m(x)) + (x^2 - m(m+1))j_m(x) j_m(x)$, $P_n$ are   Legendre polynomials, $j_m$ are   spherical Bessel functions, $h_m$ are spherical Hankel functions of the first kind and
\be
\scatZs^m (\s_j) = \frac{q_j j_m' (k a_j) j_m ( \edit{k_j} a_j) - j_m (k a_j) j_m' (\edit{k_j} a_j) }{q_j h_m'(k a_j) j_m(\edit{k_j} a_j) - h_m(k a_j) j_m '(\edit{k_j} a_j)} = \scatZs^{-m}(\s_j),
\label{eqn:Zsm}
\en
with $q_j = (\rho_j c_j)/(\rho c)$, where $\mathcal G$ is a Gaunt coefficient \cite[Eq. (A.5)]{linton_multiple_2006} defined here as
\begin{equation}
   \mathcal G(n,0;p,0;q)= \frac{\sqrt{(2n+1)(2p+1)(2q+1)}}{2 \sqrt{4 \pi}}   \int_0^\pi  P_n(\cos \theta) P_p(\cos \theta) P_q(\cos \theta) \sin \theta d \theta.
  \label{eqn:G}
\end{equation}

\section{Calculating the effective equations \eqref{eqn:AmInc} and \eqref{eqn:AmT}}
\label{sec:Integrals}
In this Appendix, we provide an indepth outline of the derivation for \eqref{eqn:AmInc} and \eqref{eqn:AmT}, which are given in terms of the unknowns $\A n(\mathbf x_2,\mathbf s)$ and $k_\eff$. The approach  here is similar to Section IV in \cite{linton_multiple_2005}.

We begin by substituting \eqref{eqn:AnsatzA1} and \eqref{eqn:AnsatzA} into the integral in the governing equation~\eqref{eqn:ensemAsystem2} which for $x_1 > {\bar x} + a_{21}$ takes the form
\begin{multline}
  \int_{\stackrel{x_2>0}{R_{2 1} > a_{21}}} \A n(\mathbf x_2,\mathbf s) \Phi_{n-m}(k R_{21},\Theta_{21}) d \mathbf x_2
    =
       \ee^{\ii \beta y_1} \ee^{-\ii (n-m) \pi } \int_{0}^{{\bar x}}  \A n(x_2,0,\mathbf s) L_{n-m} (x_2 - x_1) d  x_2
\\
+ \ii^n \ee^{-\ii n \theta_\eff} \ee^{-\ii (n-m) \pi }\A n_\eff \ee^{\ii \mathbf k_\eff \cdot \mathbf x_1} \int_{\stackrel{X>{\bar x}-x_1}{R> a_{21}}} \ee^{\ii \mathbf k_\eff \cdot \mathbf X}  \Phi_{n-m}(k R,\Theta)  d \mathbf X,
\label{eqn:IntegrateAnsatz}
\end{multline}
where $X = x_2 - x_1$ and $Y = y_2 - y_1$, so that $(R,\Theta)$ are the polar coordinates of $\mathbf X = (X,Y)$, where $R = R_{21}$ and $\Theta = \Theta_{21} -\pi$, and we define
\begin{equation}
\Phi_{n-m}(k R,\Theta) := \ee^{\ii (n-m) \Theta} H_{n-m}(k R), \quad
L_{n-m}(X) := \int_{-\infty}^\infty \ee^{\ii \beta Y}  \Phi_{n-m}(k R,\Theta) d Y.
\label{eqn:PhiL}
\end{equation}
From  \cite[Eq. (37)]{martin_multiple_2011} we have
\begin{gather}
L_n(X) =
    \begin{cases}
      \frac{2}{\alpha} (-\ii)^{n} \ee^{-\ii n\theta_\inc} \ee^{\ii \alpha X}, & \text{for} \quad X>0,
       \\
       \frac{2}{\alpha} \ii^{n} \ee^{\ii n\theta_\inc} \ee^{-\ii \alpha X}, & \text{for} \quad X<0.
    \end{cases}
\label{eqn:L}
\end{gather}

To calculate the last integral in~\eqref{eqn:IntegrateAnsatz}, it is necessary that $k_\eff \neq k$, as  $k_\eff = k$ leads to a divergent integrand. Assuming $k_\eff \not = k$ we observe that
\begin{equation*}
  \ee^{\ii \mathbf k_\eff \cdot \mathbf X} \Phi_{n-m}(k R,\Theta) =  \frac{1}{k^2 - k_\eff^2}\left[\Phi_{n-m}(k R,\Theta)\nabla^2 \ee^{\ii \mathbf k_\eff \cdot \mathbf X} - \ee^{\ii \mathbf k_\eff \cdot \mathbf X} \nabla^2 \Phi_{n-m}(k R,\Theta)\right],
\end{equation*}
because $\nabla^2 \Phi_{n-m}(k R,\Theta)  = - k^2 \Phi_{n-m}(k R,\Theta)$ and $\nabla^2 \ee^{\ii \mathbf k_\eff \cdot \mathbf X}  = - k^2_\eff \ee^{\ii \mathbf k_\eff \cdot \mathbf X}$. Then by Green's second identity we obtain
\begin{multline} \label{eqn:reducedaa1}
  \int_{\stackrel{X>{\bar x}-x_1}{R> a_{21}}}
  \left[
     \Phi_{n-m}(k R,\Theta)\nabla^2 \ee^{\ii \mathbf k_\eff \cdot \mathbf X} - \ee^{\ii \mathbf k_\eff \cdot \mathbf X} \nabla^2 \Phi_{n-m}(k R,\Theta)
  \right] d \mathbf X
  \\ =
  \int_{\partial \mathcal B}
 \left[
   \Phi_{n-m}(k R,\Theta)\frac{\partial \ee^{\ii \mathbf k_\eff \cdot \mathbf X}}{\partial \mathbf n} - \ee^{\ii \mathbf k_\eff \cdot \mathbf X} \frac{\partial \Phi_{n-m}(k R,\Theta)}{\partial \mathbf n}
 \right ] d s,
\end{multline}
where $\mathbf n$ is the outwards pointing unit normal. For $x_1 > a_{21}$, the region $\mathcal B$ is given by $X>{\bar x}-x_1$ and $R>a_{21}$, so that the boundary $\partial \mathcal B$ is the  circle $R = a_{21}$ and the line $X = {\bar x}-x_1$. The integral on the  left-hand side of \eqref{eqn:reducedaa1} reduces to   the form
\begin{equation}
  \int_{\mathcal B} \ee^{\ii \mathbf k_\eff \cdot \mathbf X} \Phi_{n-m}(k R,\Theta) d \mathbf X = \frac{1}{\alpha^2-\alpha^2_\eff}(M_\circ + M_{-}),
  \label{eqn:McircAndline}
\end{equation}
where  $k^2 - k_\eff^2 = \alpha^2 - \alpha^2_\eff$. Here, we have   \cite[Eq. (68)]{linton_multiple_2005}:
\begin{align}
  & M_\circ =
   2 \pi \ii^{n-m} \ee^{\ii(n-m)\theta_\eff} \mathcal N_{n-m}(k a_{12}, k_\eff a_{12}),
  \label{eqn:Mcirc}
\end{align}
where $\mathcal N_{n-m}$ is defined by~\eqref{eqn:Nn}, and 
\begin{multline}
  M_{-} =  -\int_{-\infty}^{\infty}\left[
    \Phi_{n-m}(k R,\Theta) \frac{\partial \ee^{\ii \mathbf k_\eff \cdot \mathbf X }}{\partial X}  - \ee^{\ii \mathbf k_\eff \cdot \mathbf X} \frac{\partial \Phi_{n-m}(k R,\Theta)}{\partial X}
  \right ] d Y 
 \\
=
\begin{cases}
  2 \frac{\alpha_\eff +\alpha}{\alpha} \ii^{n-m-1} \ee^{\ii (\alpha_\eff-\alpha)X} \ee^{\ii (n-m) \theta_\inc},
    \quad \text{for} \quad X<0,
   \\
   2 \frac{\alpha_\eff - \alpha}{\alpha} \ii^{-(n-m)-1} \ee^{\ii (\alpha_\eff+\alpha)X} \ee^{-\ii (n-m) \theta_\inc},
    \quad \text{for} \quad X>0,
\end{cases}
\label{eqn:Mline}
\end{multline}
for $X={\bar x} -x_1 <0$, which is identical to  \cite[Eq. (67)]{linton_multiple_2005} (with the replacements $\ell \mapsto{\bar x}$ and $\lambda \mapsto \alpha_\eff$, and where we have included the case $X>0$ for future reference).

From (\ref{eqn:McircAndline}) it follows that \eqref{eqn:IntegrateAnsatz} now becomes
\begin{gather*}
    \int_{\stackrel{x_2>0}{R_{2 1} > a_{21}}} \A n(\mathbf x_2,\mathbf s) \Phi_{n-m}(k R_{21},\Theta_{21}) d \mathbf x_2
  = \ii^{m} \ee^{\ii \beta y_1 }
   \left [
    \ee^{-\ii m \theta_\eff}\ee^{\ii \alpha_\eff x_1}\mathcal B^{n,m}
    +\ee^{\ii \alpha x_1}\ee^{\ii (n-m) \theta_\inc} \mathcal C^{n,m}
  \right ],
\end{gather*}
where
\begin{align}
  &\mathcal B^{n,m} =\A n_\eff \frac{2 \pi }{k^2 - k^2_\eff} \mathcal N_{n-m}(k a_{12}, k_\eff a_{12}),
  \label{eqn:Bnm}
\\
  &\mathcal C^{n,m} =  \frac{2}{\alpha}\int_{0}^{{\bar x}} (-\ii)^n \A n(x_2,0,\mathbf s)  \ee^{-\ii \alpha x_2}d  x_2 +  \frac{2}{\alpha} \ii \A n_\eff \ee^{-\ii n \theta_\eff}  \frac{\ee^{\ii (\alpha_\eff-\alpha){\bar x}}}{\alpha_\eff-\alpha}.
\end{align}
Substituting the above into~\eqref{eqn:ensemAsystem2}, assuming $x_1> \bar x + a_{21}$, and canceling common factors, we arrive at
\begin{multline}
    \ee^{\ii \alpha_\eff x_1}\ee^{-\ii m \theta_\eff} \left [\A m_\eff+
    \nfrac {} \sum_{n=-\infty}^\infty \int_\regS
     \mathcal B^{n,m}
     d\s_2^n
  \right]
\\
  +
    \ee^{\ii \alpha x_1}\ee^{-\ii m \theta_\inc} \left [ \nfrac {}  \sum_{n=-\infty}^\infty \ee^{\ii n \theta_\inc} \int_\regS
   \mathcal C^{n,m}
    d\s_2^n + 1
  \right]
   = 0,
  \label{eqn:IntegratedAsystem}
\end{multline}
having also used $I_1 = \ee^{\ii (\alpha_\eff x_1 + \beta y_1)}$. As the above must hold for all $x_1$, we can   equate the terms in the brackets to zero, which leads to \eqref{eqn:AmInc} and \eqref{eqn:AmT}.

\subsection{Effective wave for any pair distribution}
\label{sec:PairDistribution}

To calculate the effective wave for any pair distribution function $\chi$ satisfying~\eqref{eqn:pairdist_assumption}, we substitute~\eqref{eqn:PcondSGeneral} into \eqref{eqn:ensemAsystem2}, which leads to an extra integral appearing on the left side of \eqref{eqn:IntegrateAnsatz}:
\begin{multline*}
  \int_{\stackrel{\reginf}{R_{2 1} > a_{21}}}  \A n (\mathbf x_2, \s_2) \Phi_{n-m}(k R_{21},\Theta_{21}) \chi(R_{21} | \mathbf s_1, \mathbf s_2)  d \mathbf x_2 =
\\
     \ii^n \A n_\eff(\s_2) \ee^{- \ii n \theta_\eff} \ee^{-\ii (n-m) \pi} \ee^{\ii \mathbf k_\eff \cdot \mathbf x_1} \int_{\stackrel{X> \bar x -x_1}{a_{21}< R< \bar a_{21}}} \ee^{\ii
  \mathbf k_\eff \cdot \mathbf X} \Phi_{n-m}(k R,\Theta) \chi(R | \s_1, \s_2)  d \mathbf X
\\
   +\ee^{-\ii (n-m) \pi}\int_{\stackrel{-x_1\leq X \leq \bar x -x_1 }{ a_{21}< R< \bar a_{21}}} \A n (\mathbf x_2, \s_2) \Phi_{n-m}(k R,\Theta) \chi(R | \mathbf s_1, \mathbf s_2)  d \mathbf x_2,
\end{multline*}
where $X = x_2 - x_1$ and $Y = y_2 - y_1$, so that $(R,\Theta)$ are the polar coordinates of $\mathbf X = (X,Y)$, where $R = R_{21}$, $\Theta = \Theta_{21} -\pi$. The second integral on the right is zero when $x_1 > \bar x + \bar a_{21}$ because then
\[
-x_1 \leq X \leq \bar x -x_1 < - \bar a_{21},
\]
and $\mathbf X$ can not satisfy both $X < - \bar a_{21}$ and $ a_{21}< R< \bar a_{21}$.

For $x_1 > \bar x + \bar a_{21}$, the integral over the regions $X > \bar x - x_1$ and $a_{21}< R< \bar a_{21}$ reduces to the just $a_{21}< R< \bar a_{21}$, so leaving out the factors on the left, and using $\ee^{\ii \mathbf k_\eff \cdot \mathbf X} = \ee^{\ii k_\eff R \cos(\Theta-\theta_\eff)}$,
the integral becomes
\begin{multline}
 \int_{a_{21}< R < \bar a_{21}} \int_0^{2 \pi}   \ee^{\ii k_\eff R \cos(\Theta-\theta_\eff)} \ee^{\ii (n-m) \Theta} H_{n-m}(k R) \chi(R | \mathbf s_1, \mathbf s_2) R d \Theta d R,
    \\
   =
  2 \pi \ii^{n-m} \ee^{\ii (n-m) \theta_\eff } \mathcal X_\eff,
\end{multline}
where we used the Jacobi-Anger expansion and $\mathcal X_\eff$ is defined by \eqref{eqn:X_eff}. In conclusion, we should sum $ 2 \pi \A n_\eff(\s_2) \mathcal X_\eff$ to $\mathcal B_{n,m}$, appearing in equations~\eqref{eqn:Bnm} and \eqref{eqn:IntegratedAsystem}.

\subsection {Low number fraction}
\label{sec:app_LowNumberFraction}
To calculate $k_\eff^2$ we need only equation~\eqref{eqn:AmT}, which after substituting~\eqref{eqns:nfracExpansion} and
\begin{align}
  &\mathcal N_m(k a_{12},k_\eff a_{12}) \sim \frac{2 \ii}{\pi} + \nfrac {} K_{\eff 1}\frac{a_{21^2}}{2} d_m(k a_{21}),
\end{align}
with $d_m$ defined by~\eqref{eqn:d}, and then equating terms of order $\mathcal O(1)$ and  $\mathcal O(\nfrac {})$, leads to
\begin{align}
  &\A m_{\eff 0}(\s_1)  = \frac{4 \ii}{K_{\eff 1}} \sum_{n=-\infty}^\infty \int_{\regS} \A n_{\eff 0}(\s_2) d \s_2^n,
  \label{eqn:K1}
\\
  &\A m_{\eff 1}(\s_1) - \pi \sum_{n=-\infty}^\infty \int_{\regS} a_{21}^2 d_{n-m}(k a_{21}) \A n_{\eff 0}(\s_2) d \s_2^n
\notag \\
  & \quad = \frac{4 \ii}{K_{\eff 1}}  \sum_{n=-\infty}^\infty \int_{\regS}
  \left[\A n_{\eff 1}(\s_2) - \A n_{\eff 0}(\s_2) \frac{K_{\eff 2}}{K_{\eff 1}} \right]d \s_2^n.
  \label{eqn:K2}
\end{align}

Turning to \eqref{eqn:K1}, we see that $\A m_{\eff 0}(\s_1)$ is independent of both $m$ and $\s_1$.
Let $\Ab_{\eff 0} := \A n_{\eff 0}(\s_2) = \A m_{\eff 0}(\s_1) $, then we can divide both sides of \eqref{eqn:K1} by $\Ab_{\eff 0}$ (assuming $\Ab_{\eff 0} \not = 0$), and use~\eqref{eqns:FarFields} to arrive at
\begin{equation}
  K_{\eff 1} = - 4 \ii \ensem{f_\circ}(0).
  \label{eqn:solK1}
\end{equation}

Turning to \eqref{eqn:K2}, we see that
\begin{equation}
  F_{\eff} = \A m_{\eff 1}(\s_1) - \pi \Ab_{\eff 0} \sum_{n=-\infty}^\infty \int_{\regS} a_{21}^2 d_{n-m}(k a_{21}) d \s_2^n,
  \label{eqn:independent-FT}
\end{equation}
 is independent of both $m$ and $\s_1$, which
 we use to write \eqref{eqn:K2} as
\begin{align*}
   F_{\eff} &= -\frac{1}{\ensem{f_\circ}(0)}  \sum_{n=-\infty}^\infty \int_{\regS}
  \A n_{\eff 1}(\s_2) d \s_2^n
  -\ii \Ab_{\eff 0} \frac{K_{\eff 2}}{ 4 \ensem{f_\circ}(0)}
\notag \\
  & = -\frac{F_{\eff}}{\ensem{f_\circ}(0)} \sum_{n=-\infty}^\infty \int_{\regS} d \s_2^n
 -\ii \Ab_{\eff 0} \frac{K_{\eff 2}}{ 4 \ensem{f_\circ}(0)}
 + \Ab_{\eff 0}\frac{\ensem{f_{\circ\circ}}(0) }{\ensem{f_\circ}(0)},
\end{align*}
where we used~\eqref{eqn:MST-pattern}. This simplifies to
$
  K_{\eff 2}
  =
  - 4 \ii \ensem{f_{\circ\circ}}(0),
$
which together with~(\ref{eqn:solK1},\ref{eqns:nfracExpansion}) leads to the effective wavenumber \eqref{eqn:SmallNfrac}.


\bibliographystyle{RS}
\bibliography{online,Library,manual-library}



\end{document}